\RequirePackage{rotating}

\documentclass[utf8]{frontiersSCNS} 

\usepackage{url,hyperref,lineno,microtype,subcaption}
\usepackage{graphicx}
\usepackage{amssymb}

\usepackage{wrapfig}
\usepackage{tikz}
\usepackage{pgf}
\usepackage{float}
\usepackage{subcaption}
\usepackage{xcolor}
\usepackage{multirow}
\usepackage{rotating}
\usepackage{array, makecell}
\usepackage[acronym]{glossaries}
\newacronym{lrp}{LRP}{Layer-wise Relevance Propagation}
\newacronym{ad}{AD}{Alzheimer's disease}
\newcommand{\lrpfull}{\glsreset{lrp}\gls*{lrp}}

\def\keyFont{\fontsize{8}{11}\helveticabold }
\def\firstAuthorLast{Böhle {et~al.}} 
\def\Authors{Moritz Böhle\,$^{1\dagger}$, Fabian Eitel\,$^{1\dagger}$, Martin Weygandt\,$^{2}$ and Kerstin Ritter\,$^{1,*}$, for the  Alzheimer’s  Disease  Neuroimaging  Initiative\, $^{+}$} 

\def\Address{$^{1}$Charit\'e – Universit\"atsmedizin Berlin, corporate member of Freie
Universit\"at Berlin, Humboldt-Universit\"at zu Berlin, and Berlin Institute of
Health; Department of Psychiatry and Psychotherapy, Bernstein Center for Computational Neuroscience; 10117 Berlin, Germany \\
$^{2}$Charité – Universitätsmedizin Berlin, corporate member of Freie Universität Berlin, Humboldt-Universität zu Berlin, and Berlin Institute of Health; Excellence Cluster NeuroCure; 10117 Berlin, Germany.\\
$^{+}$Data  used  in  preparation  of  this  article  were  obtained  from  the  Alzheimer’s  Disease Neuroimaging  Initiative  (ADNI)  database  (adni.loni.usc.edu). As  such,  the  investigators within the ADNI contributed to the design and implementation of ADNI and/or provided data but  did  not  participate  in  analysis  or  writing  of  this  report.  A  complete  listing  of  ADNI investigators can be found at: \url{http://adni.loni.usc.edu/wp-content/uploads/how_to_apply/ADNI_Acknowledgement_List.pdf}\\
$^{\dagger}$ Equal contribution
}

\def\corrAuthor{Kerstin Ritter; Charitéplatz 1, 10117 Berlin, Germany}

\def\corrEmail{kerstin.ritter@charite.de}

\begin{document}

\onecolumn
\firstpage{1}

\title[LRP for MRI-based AD classification]{Layer-wise relevance propagation for explaining deep neural network decisions in MRI-based Alzheimer's disease classification}





\author[\firstAuthorLast ]{\Authors} 
\address{} 
\correspondance{} 

\extraAuth{}

\maketitle

\begin{abstract}

\section{}
  Deep neural networks have led to state-of-the-art results in many medical imaging tasks including Alzheimer's disease (AD) detection based on structural magnetic resonance imaging (MRI) data. However, the network decisions are often perceived as being highly non-transparent making it difficult to apply these algorithms in clinical routine. In this study, we propose using layer-wise relevance propagation (LRP) to visualize convolutional neural network decisions for AD based on MRI data. Similarly to other visualization methods, LRP produces a heatmap in the input space indicating the importance / relevance of each voxel contributing to the final classification outcome. In contrast to susceptibility maps produced by guided backpropagation (``Which change in voxels would change the outcome most?"), the LRP method is able to directly highlight positive contributions to the network classification in the input space. In particular, we show that (1) the LRP method is very specific for individuals (``Why does this person have AD?") with high inter-patient variability, (2) there is very little relevance for AD in healthy controls and (3) areas that exhibit a lot of relevance correlate well with what is known from literature. 
  To quantify the latter, we compute size-corrected metrics of the summed relevance per brain area, e.g. relevance density or relevance gain. Although these metrics produce very individual `fingerprints' of relevance patterns for AD patients, a lot of importance is put on areas in the temporal lobe including the hippocampus. After discussing several limitations such as sensitivity towards the underlying model and computation parameters, we conclude that LRP might have a high potential to assist clinicians in explaining neural network decisions for diagnosing AD (and potentially other diseases) based on structural MRI data. The published version of this article can be found at \url{https://doi.org/10.3389/fnagi.2019.00194}.

\tiny
 \keyFont{ \section{Keywords:}    Alzheimer's disease, MRI, visualization, explainability, layer-wise relevance propagation, deep learning, convolutional neural networks (CNN)} 
\end{abstract}

\section{Introduction}

	In the 2018 World Alzheimer Report, it was estimated that 50 million people worldwide were suffering from dementia and this number was projected to rise to more than 152 million people until 2050.
	The most common reason for dementia is Alzheimer's disease (AD) accounting for around 60-70\,\% of dementia cases \citep{whoAlzheimer}. 
	AD is characterized by abnormal cell death, primarily in the medial temporal lobe. 
	This cell death is thought to be rooted in protein plaques and neurofibrillary tangles, which
	restrict normal neural function \citep{bondi2017alzheimer}. The resulting atrophy is visible in structural magnetic resonance imaging (MRI) data, and derived markers (such as hippocampal volume or grey matter density) have been used to diagnose AD and predict disease progression \citep{frisoni2010clinical,rathore2017review}. In the last decade, those markers have frequently been employed in machine learning settings to allow for predictions on an individual level \citep{Kloppel2008, Orru2012, Weiner2013, Ritter2015, Ritter2016}. However, those expert features usually reflect only one part of disease pathology and the combination with standard machine learning methods, such as support vector machines, do not allow for finding new and potentially unexpected hidden data characteristics that might also be important to describe a disease. 
	
	By extracting hierarchical information directly from raw or minimally processed data, deep learning approaches \citep{lecun2015deep} can help to fill a gap here and offer a great potential for improving automatic disease diagnostics. 
	One family of algorithms that perfectly lends itself to perform non-linear feature extraction from image data and their respective classification into disease categories are convolutional neural networks (CNNs), a type of (deep) neural networks optimized for image data.
	The key idea behind CNNs is inspired by the mechanism of receptive fields in the primate’s visual cortex and relates to the application of local convolutional filters for extracting regional information \citep{LeCun1995}. They typically consist of a sequence of convolutional and pooling layers which allow for summarizing key characteristics into feature maps. 
	These feature maps can then be used by a fully-connected layer or any other classifier for solving the primary supervised learning problem (e.\,g.\,AD classification).
	CNNs have been proven to be very successful in a wide range of medical imaging applications \citep{litjens2017survey}, including AD detection based on neuroimaging data (e.\,g.\,\citet{ Gupta2013, Suk2014, Payan2015,Sarraf2016,Korolev2017}; for a review, see \citet{Vieira2017}). 
	
\begin{figure} 
	\centering
	\medskip\par\par
	\medskip\par\par
	\includegraphics[clip, page=1, trim=2cm 1cm 2cm 0cm, width=.85\textwidth]{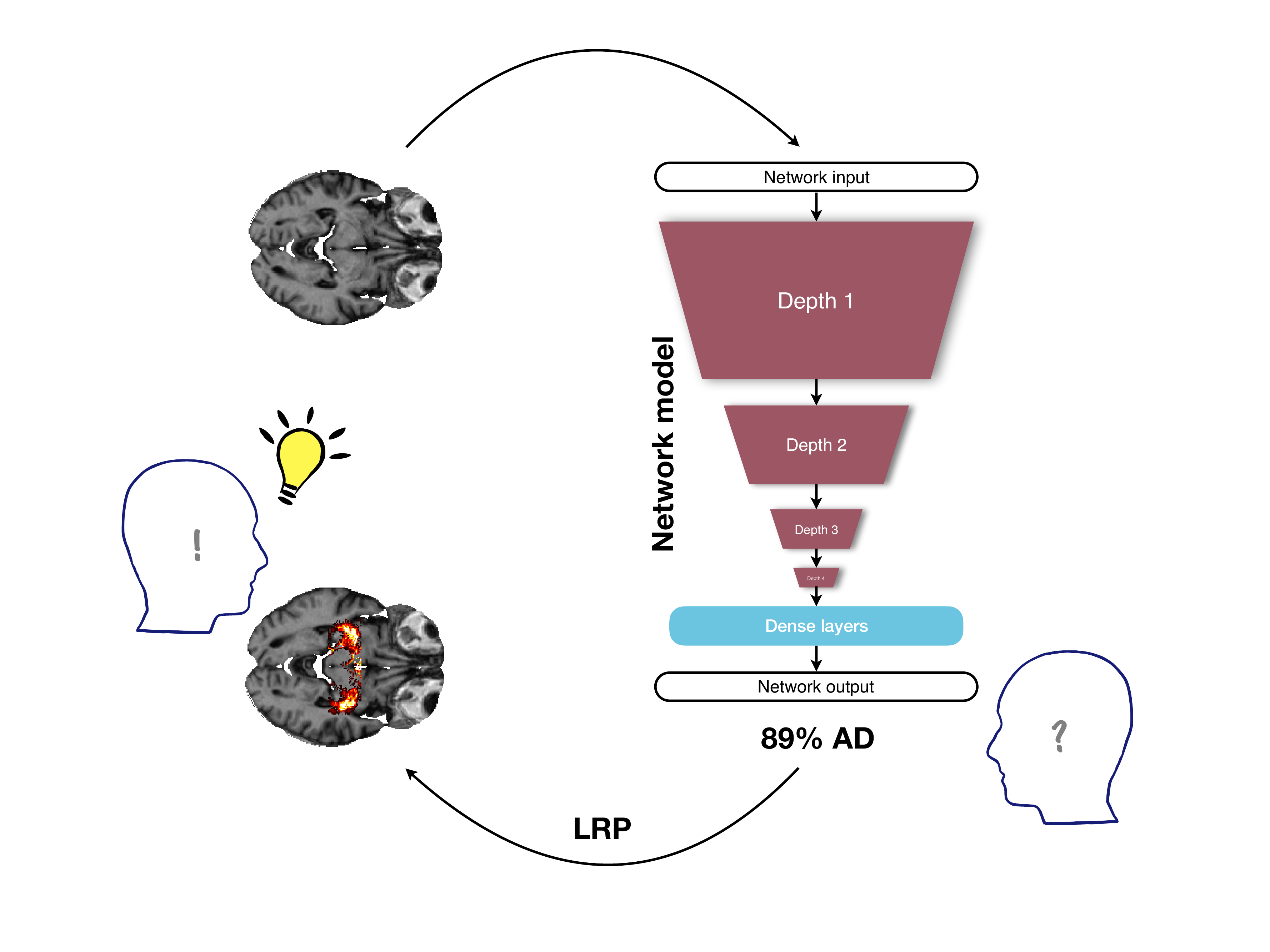}
	\caption{Illustration of the benefit of visualization in a deep learning framework for diagnosing Alzheimer's disease (AD) based on structural MRI data. Deep neural networks are often criticized for being non-transparent, since they usually provide only one single class score as output and do not explain what has led to this particular network decision; in this example, the MRI input is classified as belonging to the group of AD patients with a probability of 89\%. When no further information is given, the medical expert is not able to base any medical treatment on this number, since the underlying reasons are unclear. The layer-wise relevance propagation method (LRP) might alleviate this problem by additionally providing a heatmap in which the positive contributions to the class score (89\% AD) are highlighted. Here, the class score is supplemented by the additional information that in this particular subject AD relevance has been found in the hippocampus, an area known to be affected in AD. By providing a visual explanation, the LRP framework allows the medical expert to make a much more informed decision.}
	\label{fig:schema}
	\end{figure}

	Despite this success, automatically learning the features comes at a cost: the decisions of neural networks are notoriously hard to interpret in retrospect. Therefore, deep learning methods including CNNs often face the criticism that they are ``black-box'' \citep{Castelvecchi2016}. In contrast to some simpler learning algorithms, in particular decision trees, they do not offer a simple and comprehensible explanation; their architecture is complex and consists of several to many layers with hundreds of thousands parameters that need to be trained. In the medical domain, however, it is imperative to base diagnoses and subsequent treatments on an informed decision and not on a single yes / no answer of an algorithm. Therefore, if CNNs should support clinicians in their daily work, ways have to be found to visualize and interpret the network's `decision' (see Fig. \ref{fig:schema}). 
	In the last years, a number of suggestions have been made to visualize \textit{what} is actually learned by a CNN. Besides straightforward methods such as the extraction of activations during convolution or the visualization of weights, among the most well-known techniques for visualization are the sensitivity analysis by \citet{Simonyan2013}, guided backpropagation by \citet{springenberg2014}, the deep visualization toolbox of \citet{Yosinski2015} based on regularized optimization, and the deconvolution and occlusion method by \citet{Zeiler2014}. In Alzheimer's research only a very few studies exist that looked into such visualization methods \citep{esmaeilzadeh2018end,rieke2018visualizing,  yang2018visual}.
	
	Most promising for the use in the medical imaging domain is the generation of an individual heatmap for each patient, which lies in the same space as the input image and indicates the importance of each voxel for the final (individual) classification decision. By allowing for a human-guided, intuitive investigation of what drives the classifier to come to a certain classification decision, individual heatmaps hold great potential in assisting and understanding diagnostic decisions performed by deep neural networks.
	However, for any visualization method that produces heatmaps, it is very important to understand how they are computed and what their limitations are.
	In natural images, for example, it has been argued that methods relying on gradients (e.g. sensitivity analysis or guided backpropagation) only measure the susceptibility of the output to changes in the input and might not necessarily coincide with those areas on which the network bases its decision.
	A powerful method to overcome this limitation is layer-wise relevance propagation (LRP, \citet{10.1371/journal.pone.0130140}), which decomposes the network's output score (e.g. for AD)
	into the individual contributions of the input neurons while keeping the total amount of relevance constant across layers (conservation principle). In contrast to showing `susceptibility maps' as gradient-based methods, the heatmap does not rely on gradients, but takes into account model parameters (i.\,e.\,weights) and neuron activations \citep{10.1371/journal.pone.0130140, Samek2015}. By this, the heatmaps are less prone to group effects in the data. 
	Intuitively, LRP has the potential to answer the question ``what speaks for AD in this particular patient?" as opposed to ``which change in voxels would change the outcome most?" addressed in gradient-based approaches. In terms of explainability, LRP has been shown to be superior to those gradient methods and deconvolution methods in three natural imaging data sets \citep{Samek2015}. In cognitive neuroscience, the LRP method has been recently applied to single-trial EEG and functional MRI classification \citep{Sturm2016, Thomas2018}. To the best of our knowledge, it has so far not been applied in clinical disease classification based on structural MRI data.

	In this study, we use LRP to explain individual classification decisions for AD patients and healthy controls (HCs) based on a CNN trained on structural MRI data (T1-weighted MPRAGE) from the Alzheimer's Disease Neuroimaging Initiative (ADNI\footnote{\url{http://adni.loni.usc.edu/}}). Based on the trained CNN model, we generated LRP heatmaps for each subject in the test set. Importantly, each heatmap indicates the voxel-wise relevance for the particular classification decision (AD or HC). To spot the most relevant regions for AD classification, we computed average heatmaps across AD patients and HCs, which we then further split into correct and wrong classification decisions (i.\ e.\ true positives, false positives, true negatives, false negatives). To analyze the relevance in different brain areas according to the Scalable Brain Atlas by Neuromorphometrics Inc.\ \citep{bakker2015scalable}, we suggest size-corrected metrics and compared these metrics between LRP and guided backpropagation. We have chosen guided backpropagation as a baseline method because (1) sensitivity analysis is the most common method for generating heatmaps, (2) it results in more focused heatmaps than only using backpropagation \citep{rieke2018visualizing} and (3) it is better comparable to LRP than occlusion methods with respect to our relevance measures. On an individual level, we analyzed the heatmap patterns of single subjects (`relevance fingerprinting') and correlate them with the hippocampal volume as a key biomarker of AD. We show that the LRP heatmaps succeeded in depicting individual contributions to AD diagnosis and might hold great potential as a diagnostic tool.
	
\section{Materials and Methods}

\subsection{Data and preprocessing}
\label{sec:data}
Data  used  in  the  preparation  of  this  article  were  obtained  from  the  Alzheimer’s  Disease Neuroimaging  Initiative (ADNI, RRID:SCR\_003007) database  (adni.loni.usc.edu). The  ADNI  was  launched  in 2003 as  a  public-private  partnership,  led  by  Principal  Investigator  Michael  W.  Weiner, MD. The primary goal of ADNI has been to test whether serial magnetic resonance imaging (MRI),  positron  emission  tomography  (PET),  other  biological  markers,  and  clinical  and neuropsychological  assessment  can  be  combined  to measure  the  progression  of  mild cognitive impairment (MCI) and early Alzheimer’s disease (AD). For up-to-date information, see \url{www.adni-info.org}.\\
We included structural MRI data of all subjects with Alzheimer's disease (AD) and healthy controls (HCs) listed in the ``MRI collection - Standardized 1.5T List - Annual 2 year". 
The subjects in the data set are labelled as AD if the Clinical Dementia Rating (CDR) score \citep{morris1993clinical} was greater than 0.5. HCs are selected as those subjects with a CDR score of 0. In total, we included 969 individual scans (475 AD, 494 HC) of 193 AD patients and 151 HCs (up to three time points). 
All scans were acquired with 1.5 T scanners at various sites and had undergone gradient non-linearity, intensity inhomogeneity and phantom-based distortion correction. We downloaded T1-weighted MPRAGE scans and non-linearly registered them to the 1mm resolution 2009c version of the ICBM152 reference brain using Advanced Normalization Tools (ANTs\footnote{\url{http://stnava.github.io/ANTs/}}). 
This has been done to (1) ensure a relative alignment across subjects, (2) allow the convolutional neural network to extract more robust features, and (3) be able to analyze the heatmaps in a common space. For the region-wise analysis of heatmaps, we used the Scalable Brain Atlas by Neuromorphometrics Inc.\ \citep{bakker2015scalable} available in SPM12\footnote{\url{http://www.fil.ion.ucl.ac.uk/spm/software/spm12/}}.
A list of all areas included can be found in the SPM12 package.

\subsection{Convolutional neural network architecture}
\label{sec:visual}
    Convolutional neural networks (CNNs) are neural networks optimized for array data including images or videos \citep{lecun2015deep}. In addition to input and output layer, they consist of several hidden layers including convolutional and pooling layers. In convolutional layers, in contrast to fully-connected layers, the weights and the bias terms are shared between all neurons in a given layer for a given filter. This means that each of the neurons applies the same \textit{filter} or \textit{kernel} to the input, but at a different position, usually with a 
    displacement (often called stride) of 1-3 between neighbouring neurons. Since these filters are learned via the backpropagation algorithm, CNNs do not rely on hand-crafted features, but can be applied to minimally processed data \citep{lecun2015deep}. CNNs have been very successfully applied to a large number of applications including
    image and speech recognition \citep{Krizhevsky2012,abdel2014convolutional,long2015fully} as well as medical imaging and AD classification based on MRI data \citep{Gupta2013, Suk2014,Payan2015, Sarraf2016, Korolev2017,litjens2017survey, Vieira2017}
    
    The model in the present study consists of four convolutional blocks followed by two fully-connected layers.
	Each block features a convolutional layer with $f$ filters ($f=8,16,32,64$) and filter sizes of 3x3x3. Every convolutional layer is followed by batch normalization and max pooling with window sizes $w$x$w$x$w$ ($w = 2, 3, 2, 3$).
	The fully-connected layers contain 128 and 2 units respectively and dropout ($p=40 \%$) is applied before each. The final fully-connected layer, which is activated by a softmax function serves as the network output, providing the class scores for HCs (first unit) and AD (second unit) respectively. As an optimizer Adam \citep{Kingma2015} was used with an initial learning rate of 0.0001 and a weight decay of 0.0001.
	The data was split into a training data set (163 AD patients, 121 HCs; 797 images in total), a validation set for optimizing the hyperparameters (18 AD patients, 18 HCs; 100 images in total) and a test set (30 AD patients, 30 HCs; 172 images in total). To ensure independence between training and test data, we performed the split of the data on the level of patients instead of images.
	The data was augmented during training by flipping the images along the sagittal axis ($p=50 \%$) and translated along the sagittal axis between -2 and 2 voxels. 
   When the model did not improve for 8 epochs on the validation set, training was stopped.
   The training epoch (i.e. model checkpoint) with the best validation accuracy (91.00\%) was then applied to the test data, resulting in a classification accuracy of $87.96\%$.

\subsection{Visualization methods}	
\subsubsection{Layer-wise relevance propagation (LRP)}
	In the following, we will introduce the \lrpfull\ algorithm by \cite{10.1371/journal.pone.0130140}.
The core idea underlying the LRP algorithm for attributing relevance to individual input nodes 
	is to trace back contributions to the final output node layer by layer. 
	While several different versions of the LRP algorithm exist, they all share the same principle:
	the total relevance -- e.\,g.\,the activation strength of an output node for a certain class -- is conserved
	per layer; each of the nodes in layer $l$ that contributed to the activation of a node $j$ in the 
	subsequent layer $l+1$ gets attributed
	a certain share of the relevance $R^j_{l+1}$ of that node. Overall, the sum over the relevances of all
	nodes $i$ contributing to neuron $j$ in layer $l$ must sum to $R^j_{l+1}$, such that the total relevance per layer 
	is conserved:
	\begin{equation}
	\sum_{i} R^{\, i \rightarrow j  }_{\, l, l+1} = R^{\,j}_{\,l+1}
	\end{equation}
	 There are different ways in which the relevance can be distributed over the input nodes $i$ and 
	 different rules for how to distribute the relevances have been proposed. In this paper, we used the $\beta$-rule (as described in \citet{binder2016layer}): 
	\begin{align}
	     R^{\, i \rightarrow j  }_{\, l, l+1} = \left((1+\beta) \frac{z^+_{ij}}{z^+_{j}} - \beta \frac{z^-_{ij}}{z^-_{j}}  \right)R^{\,j}_{\,l+1} \quad \textbf{.}
	\end{align}
	 
	 Here, $z^{+/-}_{ij}$ refers to the amount of positive/negative input that node $i$ contributed to node $j$. The individual contributions are divided by the 
	sum over all positive/negative contributions of the nodes in layer $l$,  $z^{+/-}_{j} = \sum_i z^{+/-}_{ij}$, such that the relevance is conserved from layer $l+1$ to layer $l$.
	 We have chosen this rule, as it allows for adjusting how much weight is put on positive contributions relative to inhibitory contributions that benefit the AD score. 
	 LRP with a $\beta$ value of zero allows only positive contributions to be shown in the heatmap, whereas non-zero $\beta$ values additionally correct for the inhibitory effects of neuron activations.
	When diagnosing AD, the network needs to balance structural evidence speaking for and against AD. Any given local area that looks \textit{healthy} to the network, might have inhibitory effects on the AD score, as it correlates more with HC patients. As the network increases its receptive field size throughout the layers, healthy areas within this receptive field might inhibit the contribution of affected areas to the final class score of AD. By reversing this process with LRP, positive contributions lying closer to healthy areas will thus obtain a lower relevance score, as they overlap with inhibited receptive fields. This leads to sparser heatmaps, see also \citet{binderbeta}), and might disproportionately affect small structures surrounded by `healthy areas'. As AD -- especially in the early stages of the disease -- can affect brain areas in a highly localized manner, heatmaps obtained with lower $\beta$ values might therefore be more meaningful, as they highlight \textit{all} positive contributions, irrespective of their surroundings.
	Accordingly, we focus in the present study on $\beta=0$, but additionally test the robustness for varying values of $\beta$ ($\beta= 0, 0.25,0.5,0.75,1$).
	 
	 	 For a more detailed description of the LRP algorithm, we kindly refer
	 the reader to \cite{10.1371/journal.pone.0130140, Montavon2018}.
	A PyTorch implementation of the LRP algorithm has been developed for the current work and is available on github.\footnote{\url{https://www.github.com/moboehle}}

	\subsubsection{Guided backpropagation (GB)}
	In order to emphasize and point out the advantages of LRP as a diagnostic tool, we compared it to a
	gradient-based method, the guided backpropagation (GB) algorithm \citep{springenberg2014}. 
	In GB, the absolute values of the gradient of the output with 
	respect to the input nodes is shown as a heatmap, with the additional twist that negative gradients are set to zero at the rectification layers of the network. As was shown by \citet{rieke2018visualizing}, 
	`rectifying' the gradients in the backward pass leads to more focused heatmaps.
	
	\subsection{Analyzing the classification decisions}
	The CNN model was evaluated on each MR image from the test set and, subsequently, both the LRP as well as the GB algorithm were used to produce a heatmap for each MR image. In the case of LRP, we produced separate heatmaps for each $\beta$ value. We analyzed the resulting heatmaps (1) group-wise to distill those regions, which are particularly `important'  for the AD classification and (2) individually to understand the network decisions per sample and find differences between subjects. For the former, we computed an average AD heatmap (obtained from all AD subjects) and an average HC heatmap (obtained from all HCs), which we then further split into a true positive heatmap (i.\,e.\,average heatmap of clinically validated AD patients, who are classified as AD), a false positive heatmap (i.\,e.\, average heatmap of HCs classified as AD), a true negative heatmap (i.\,e.\, average heatmap of HCs classified as HC) and a false negative heatmap (i.\,e.\,average heatmap of clinically validated AD patients classified as HC). 
	For GB, these heatmaps highlight those areas to which the network is on average most susceptible.
	For LRP, they show the average relevance of each voxel for contributing to the AD score. 
    All LRP heatmaps show the average relevance for the same class (AD), 
	such that they can be compared on the same scale (relevance for AD diagnosis). As the AD scores of HCs typically range between 0 and 0.5, there will be relevance for AD in HCs, too.

\subsection{Atlas-based importance metrics}
	\label{sec:metrics}
	To quantitatively analyze the heatmaps and the underlying CNN model, we assessed 
	the importance of different brain areas 
	-- as defined by the Neuromorphometrics brain atlas \citep{bakker2015scalable} -- 
    by using the following three metrics for both LRP and GB.
	
\begin{figure*}[p]
    \vspace{0cm}
    \centering
  
  \includegraphics[clip, trim=0cm 0cm 0cm 0cm,page=1,
  width=\textwidth]{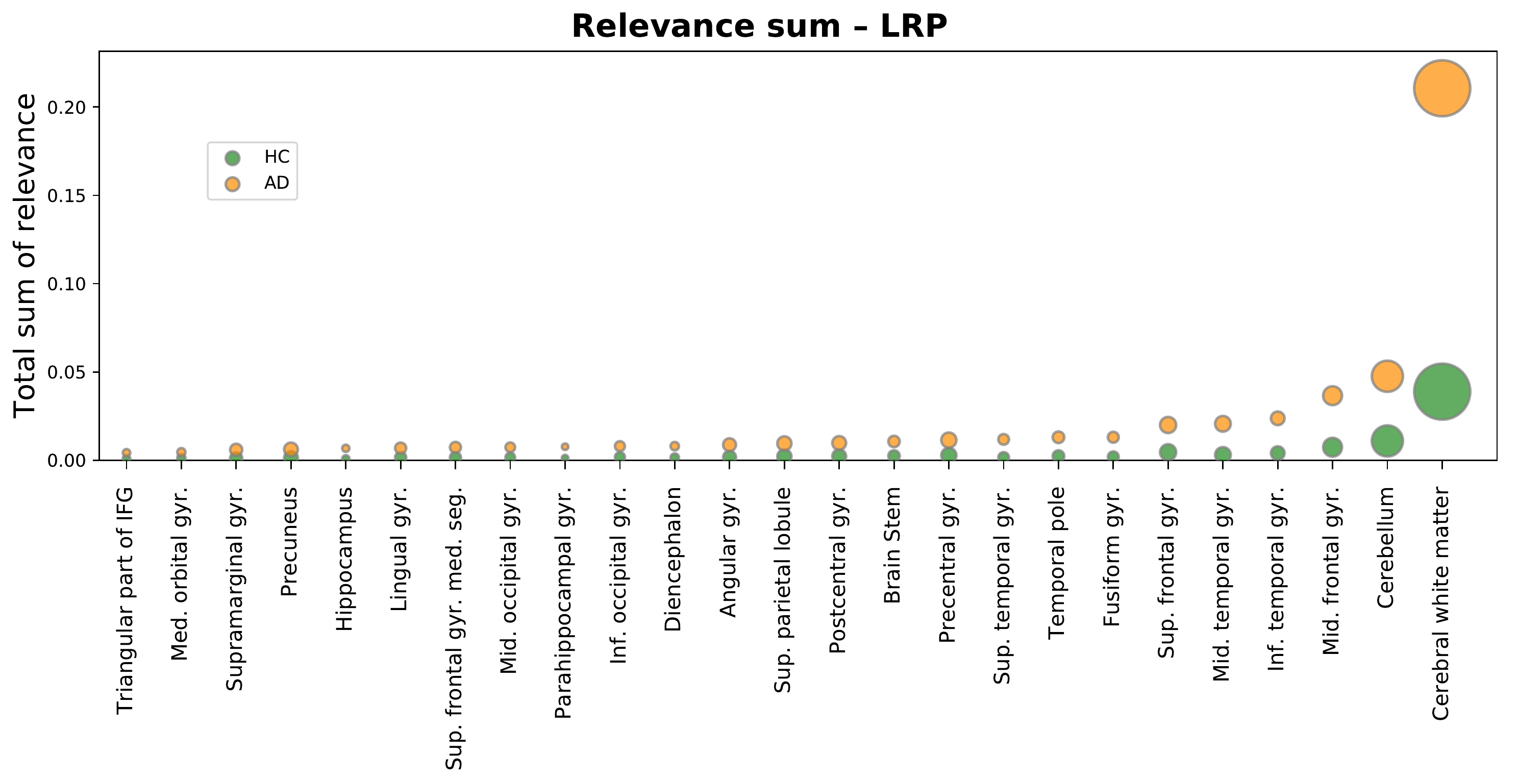}
  \includegraphics[clip, trim=0cm 0cm 0cm 0cm,page=1,
  width=\textwidth]{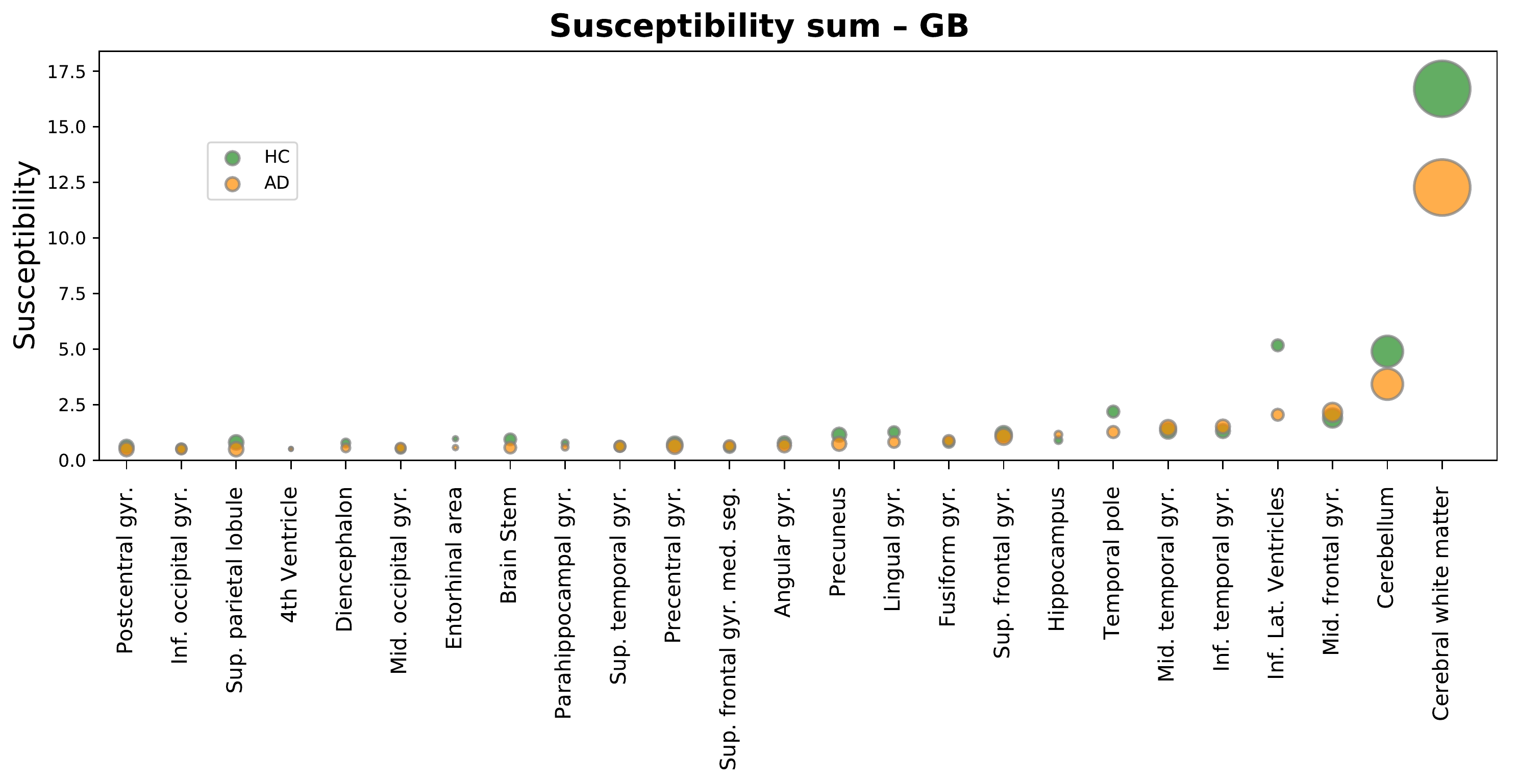}
 
    \captionof{figure}{
	Absolute sum of relevance (LRP, top) and absolute sum of susceptibility (GB, bottom) is shown for different brain areas. 
	Susceptibility refers to the absolute value of the GB gradients.
	Only the top 25 most important areas under this metric are shown for LRP and GB respectively. The circles show the average sum for each area over all AD patients (orange) and all healthy controls (HCs, green) in the test set. By setting the metric to linearly scale with the corresponding brain area size, it becomes clear that this metric is correlated with the size of the brain areas.
	}
	\label{fig:abs_evdc}
  
    \end{figure*}

	\noindent\textbf{Sum of AD importance per area.}
	As a first metric of importance, the resulting heatmap values were simply summed per area.
	While this can already be taken as a measure of importance, 
	the resulting importance scores are highly correlated to the area size, see Fig. \ref{fig:abs_evdc}. Therefore, two size-independent 
	metrics for importance were additionally analyzed in more detail: the size-normalized sum, and the average gain (ratio) when comparing to
	the average HC patient.

	\noindent	\textbf{Size-normalized AD importance metric.} 
	For diagnostic purposes, it can be particularly interesting to identify areas that
	over their entire volume carry a lot of information, i.\ e.\ areas with high \textit{relevance density} or, in GB, \textit{susceptibility density}. Therefore, we divided here the sum of AD importance per area by the size of the area (i.\,e.\,number of voxels), which corresponds to the regional mean relevance/susceptibility.
	While low values over large areas 
	might be due to statistical fluctuations in the data, clusters of 
	relevance (LRP) or susceptibility (GB) in a very confined area
	could be indicative of the presence of certain biomarkers for AD. 
		
	\noindent\textbf{Gain -- ratio of values with respect to the average HC.}
		Lastly, it is important to note that HCs are not `relevance-free' under the LRP algorithm: HCs
		might exhibit certain structural elements in their brains that are correlated with the AD diagnosis.
		While the network might still classify them as HC, these structures lead to a class score greater than
		zero for virtually every subject. Thus, as an additional metrc, we will look at the `gain' in relevance (LRP) and susceptibility (GB) per area, i.\,e.\,the ratio to the average HC in that area.
    	By doing this, those areas that differ most between the
		two cases will be attributed the highest importance.

\section{Results}
\begin{sidewaysfigure*}[p]
    
    \centering
  \begin{tabular}{@{}cccc@{}}
    \includegraphics[clip, trim=0cm 0cm 0cm 0cm,page=1,
    width=.199\textwidth]{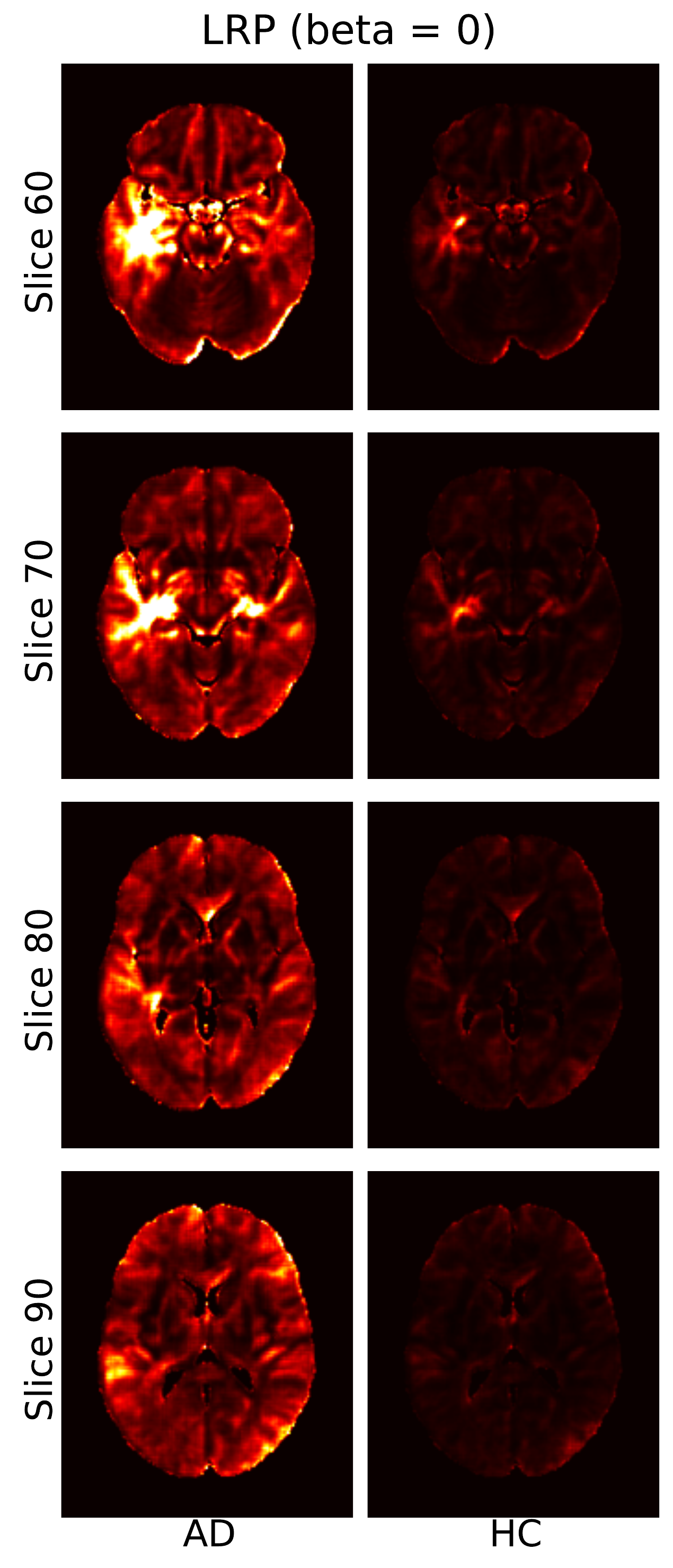} &
    \includegraphics[clip, trim=0cm 0cm 0cm 0cm,page=1,
    width=.199\textwidth]{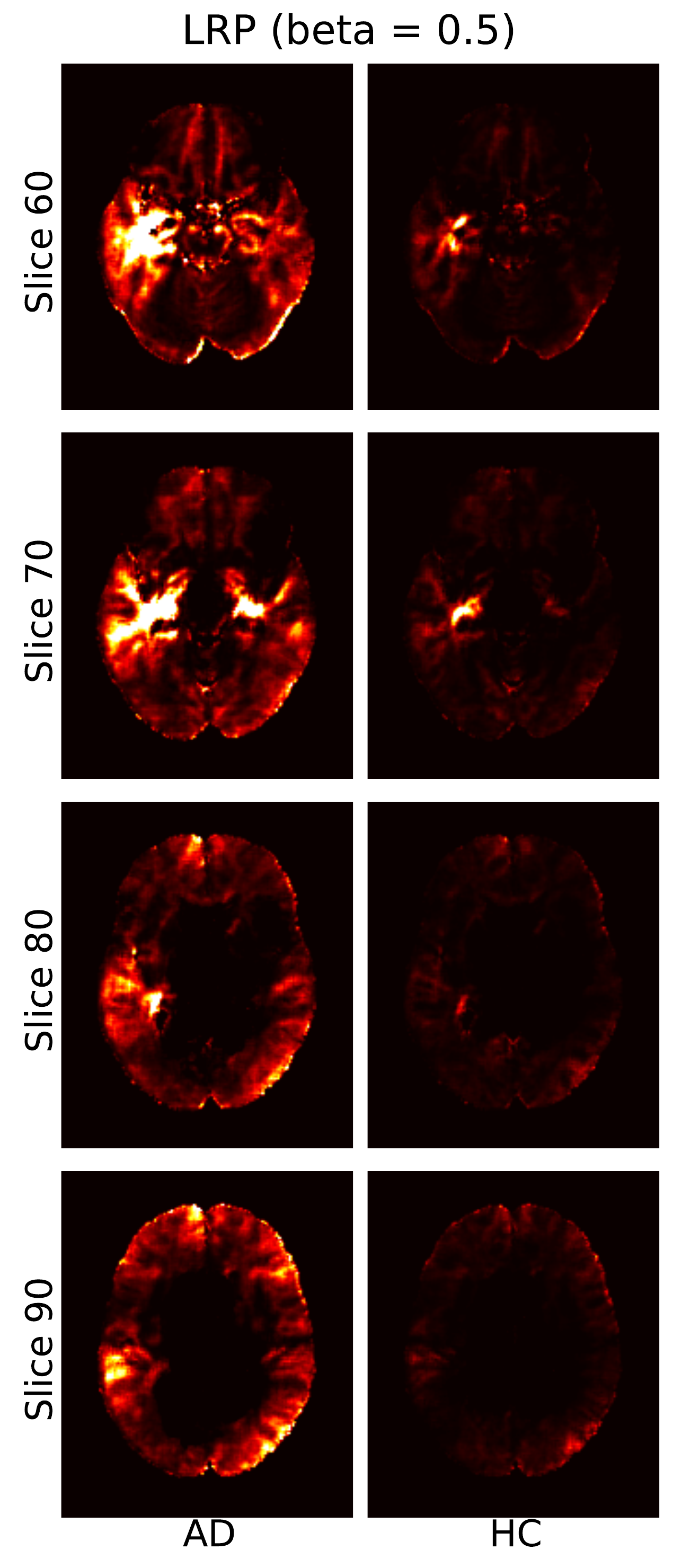} &
    \includegraphics[clip, trim=0cm 0cm 0cm 0cm,page=1,
    width=.25\textwidth]{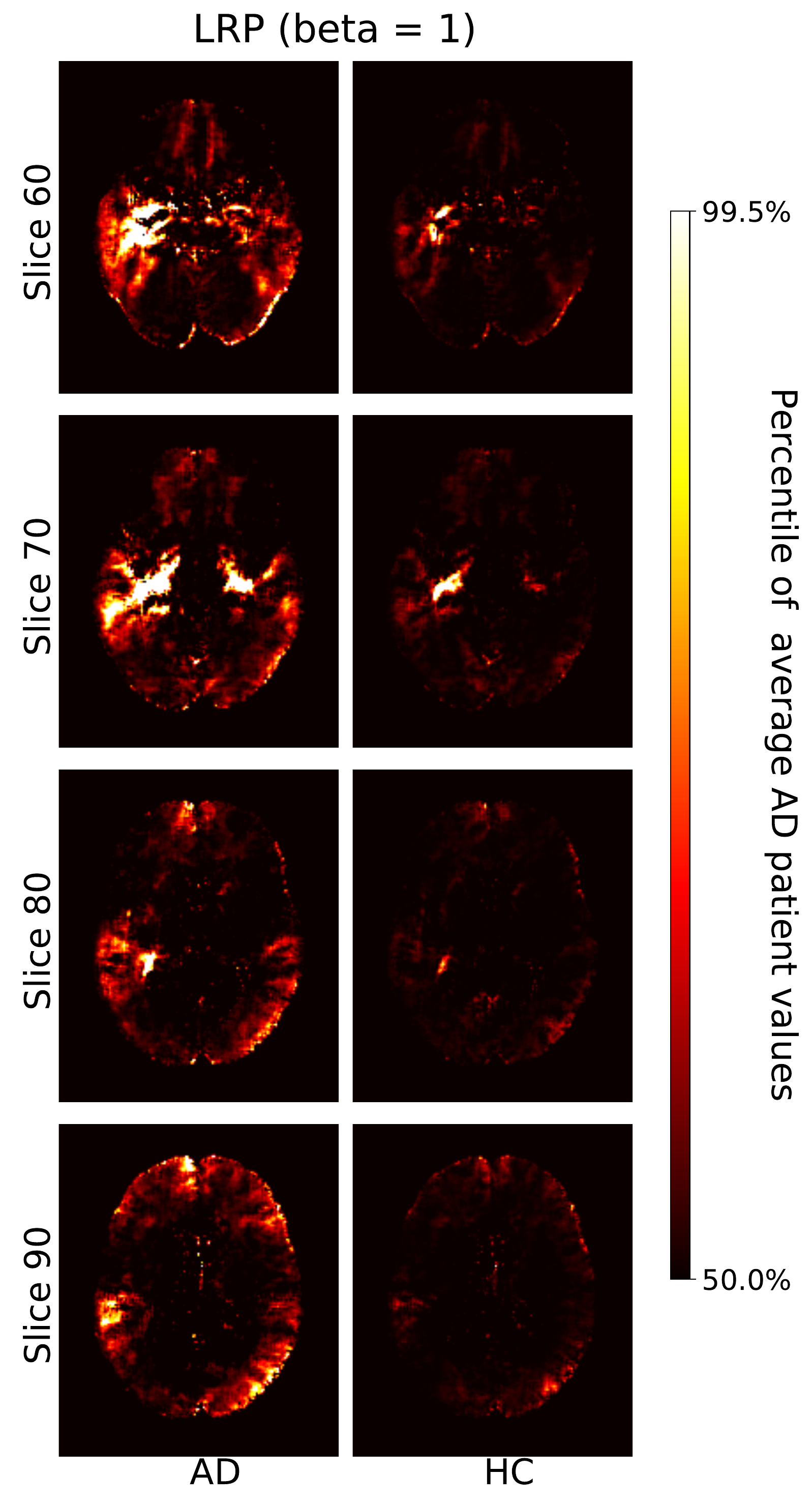} &
    \includegraphics[clip, trim=0cm 0cm 0cm 0cm,page=1,
    width=.25\textwidth]{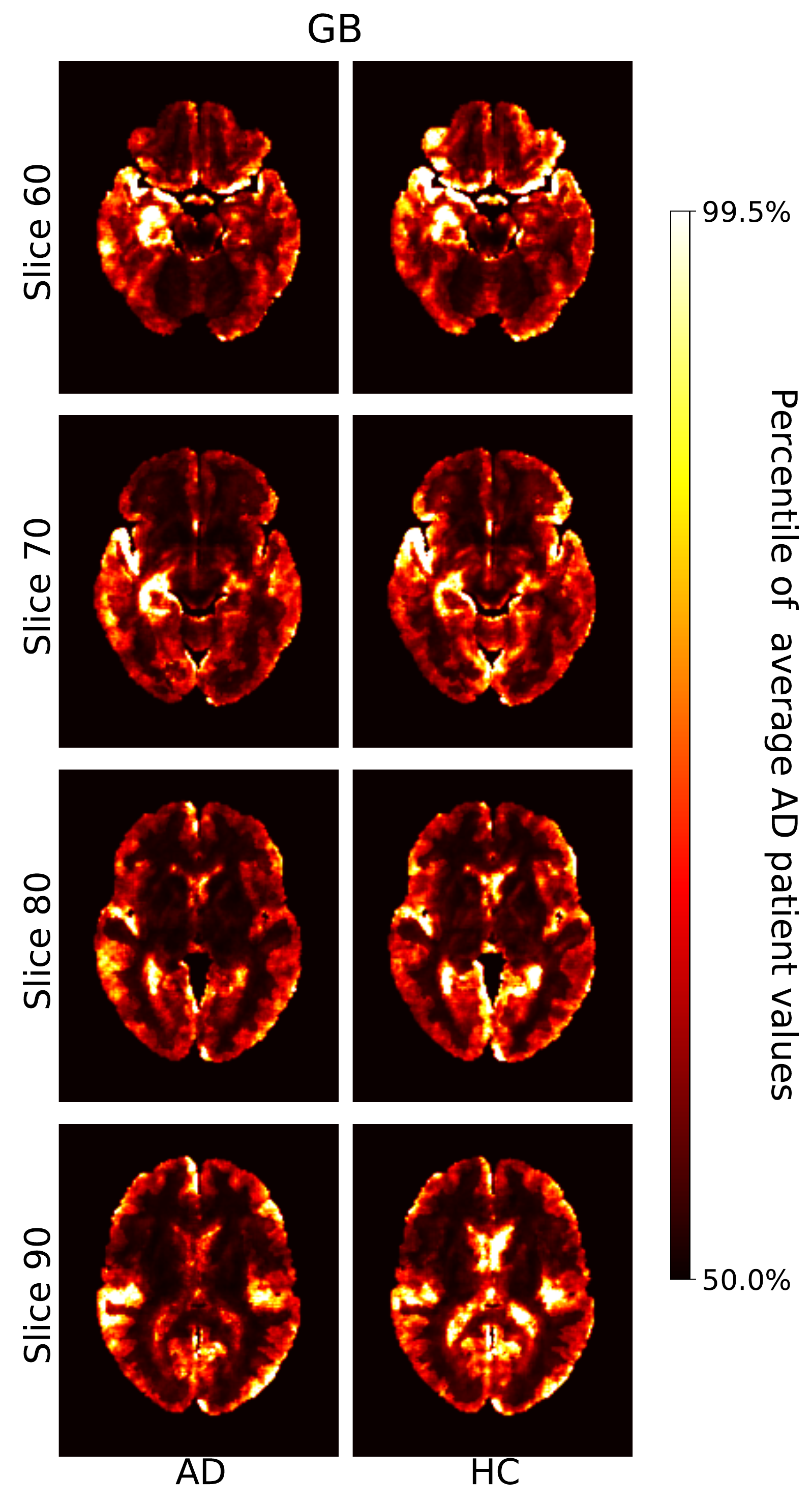}
    \end{tabular}
 
    \captionof{figure}{
	Average heatmaps for AD patients and healthy controls (HCs) in the test set are shown separately for LRP with $\beta=0,0.5,1$ (left) and GB (right).
	The scale for the heatmap is chosen relative to the average AD patient heatmap for LRP and GB respectively. Hence, values in the average heatmaps that are higher than the 50th percentile and lower than the 99.5th percentile are linearly color-coded as shown on the scale. Values below (above) these numbers are black (white).
	}
	\label{fig:LRP-GB-AD_comparison_betas}
    \end{sidewaysfigure*}

\begin{sidewaysfigure*}[p]
    
    \centering
  \begin{tabular}{@{}cc@{}}
    \includegraphics[clip, trim=0cm 0cm 0cm 0cm,page=1,
    width=.45\textwidth]{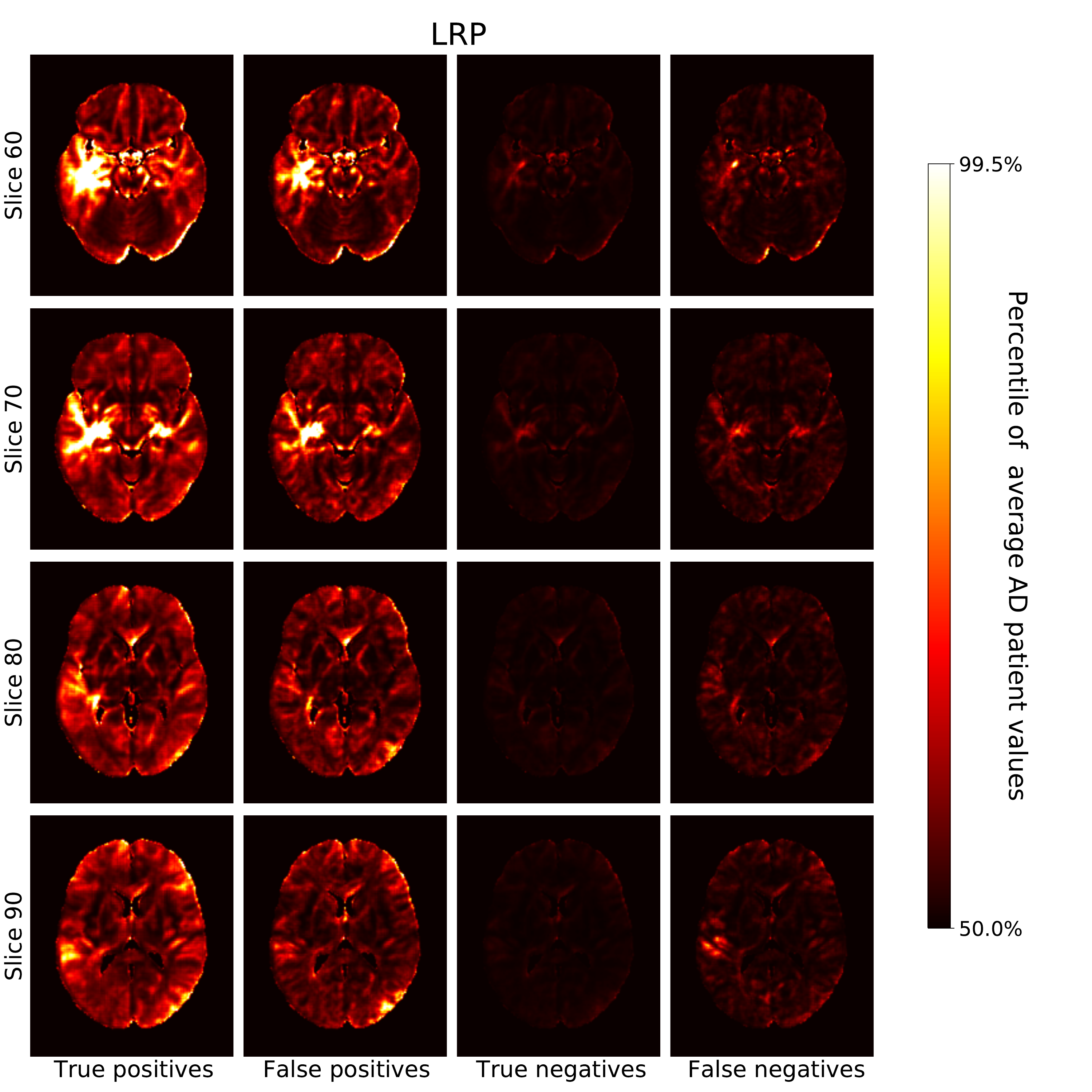} &
    \includegraphics[clip, trim=0cm 0cm 0cm 0cm,page=1,
    width=.45\textwidth]{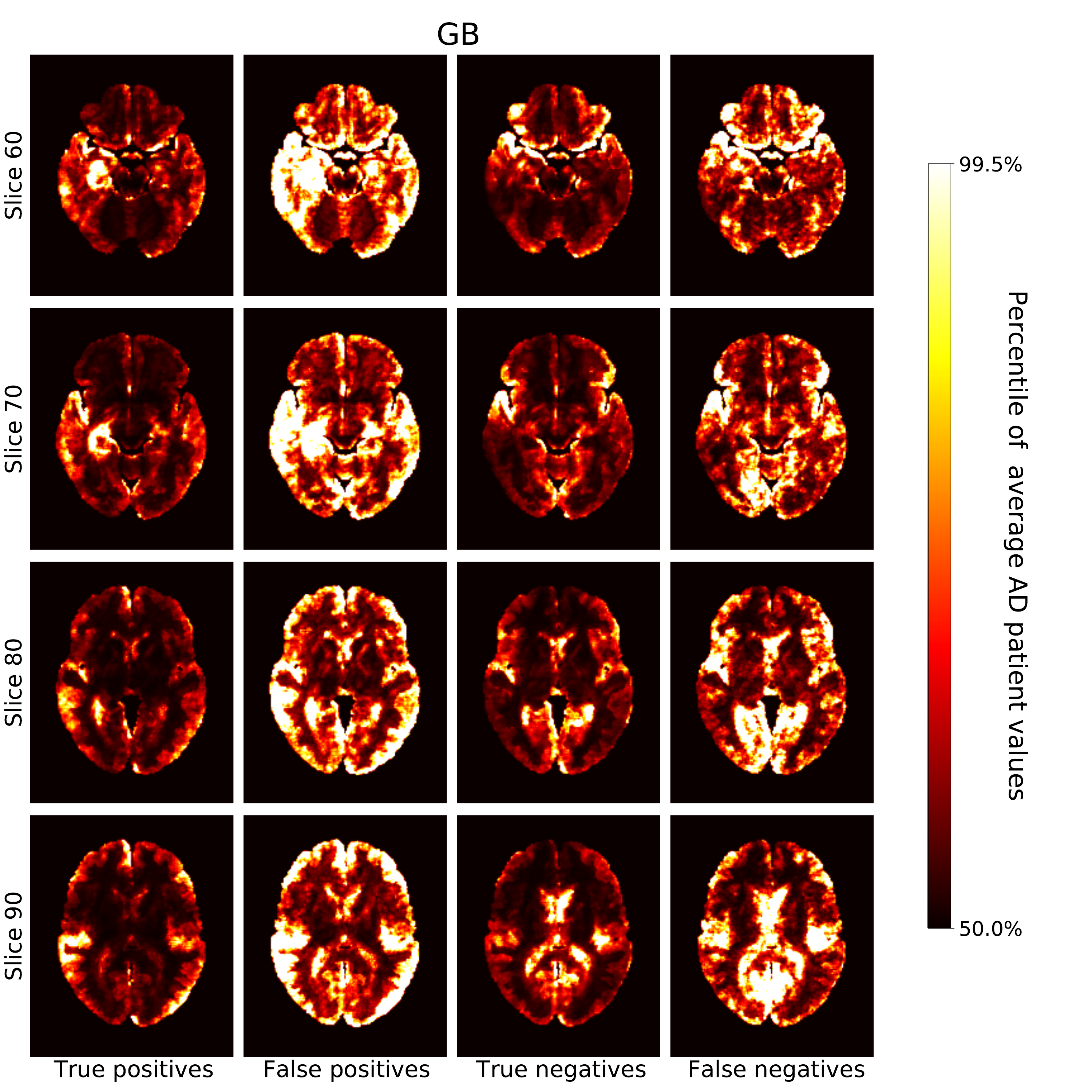}
    \end{tabular}
 
    \captionof{figure}{ 
The average heatmaps over all subjects in the test set
	are plotted for the following cases (left to right): true 
	positives, false positives, true negatives, and false negatives; separately for LRP with $\beta=0$ (left) and GB (right).
	For each heatmap, the color-coding is the same as in 
	Fig.\ \ref{fig:LRP-GB-AD_comparison_betas}, i.\ e.\ with all values
	smaller than the 50th percentile of the average AD patient in black, 
	increasing values going over red to yellow, and all values greater 
	than the $99.5$th percentile in white.
	}
	\label{fig:LRP-GB-cases}
    \end{sidewaysfigure*}

\begin{figure*}[p]

    \vspace{0cm}
    \centering
  \includegraphics[clip, trim=0cm 0cm 0cm 0cm,page=1,
  width=0.95\textwidth]{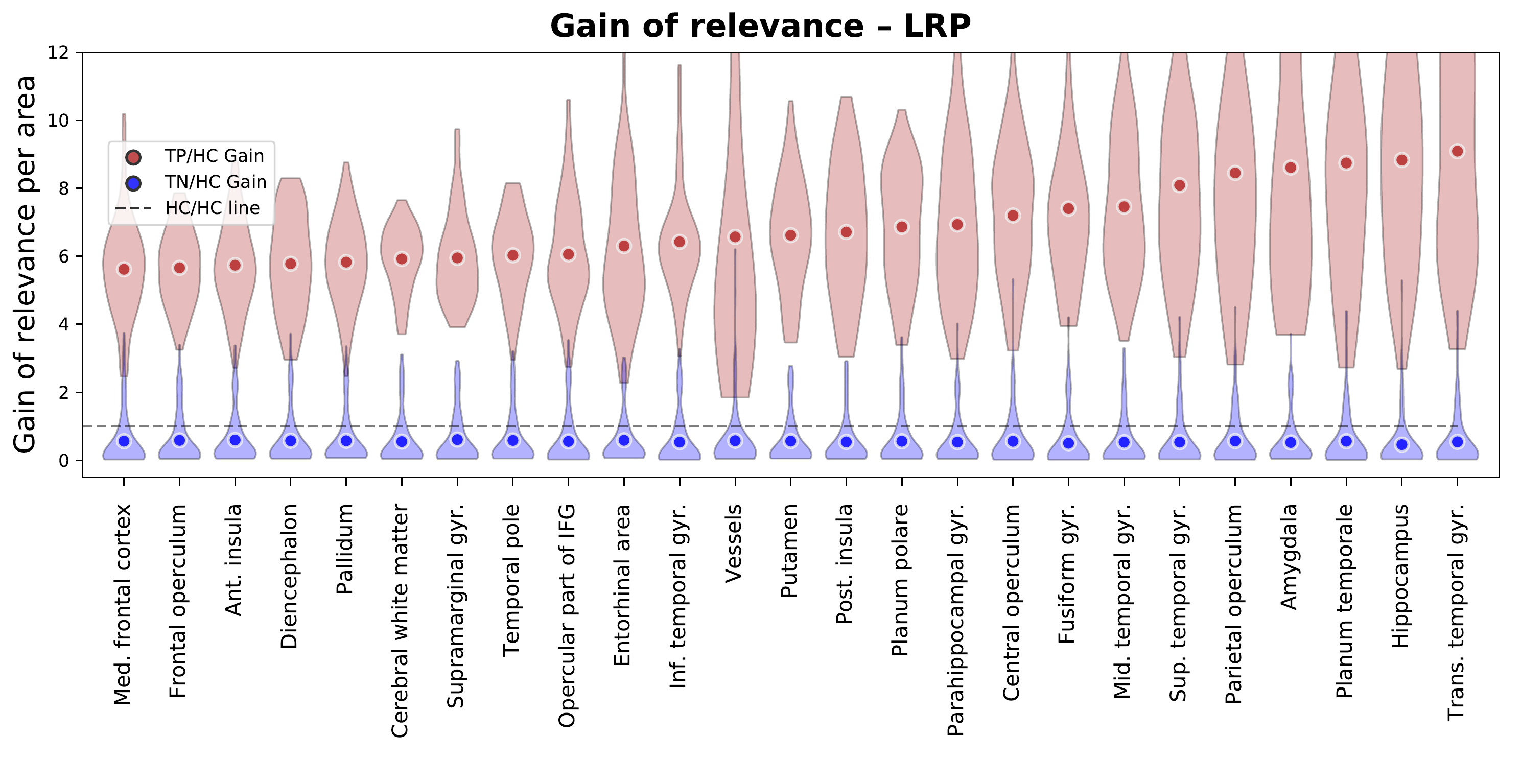}
  \includegraphics[clip, trim=0cm 0cm 0cm 0cm,page=1,
  width=0.95\textwidth]{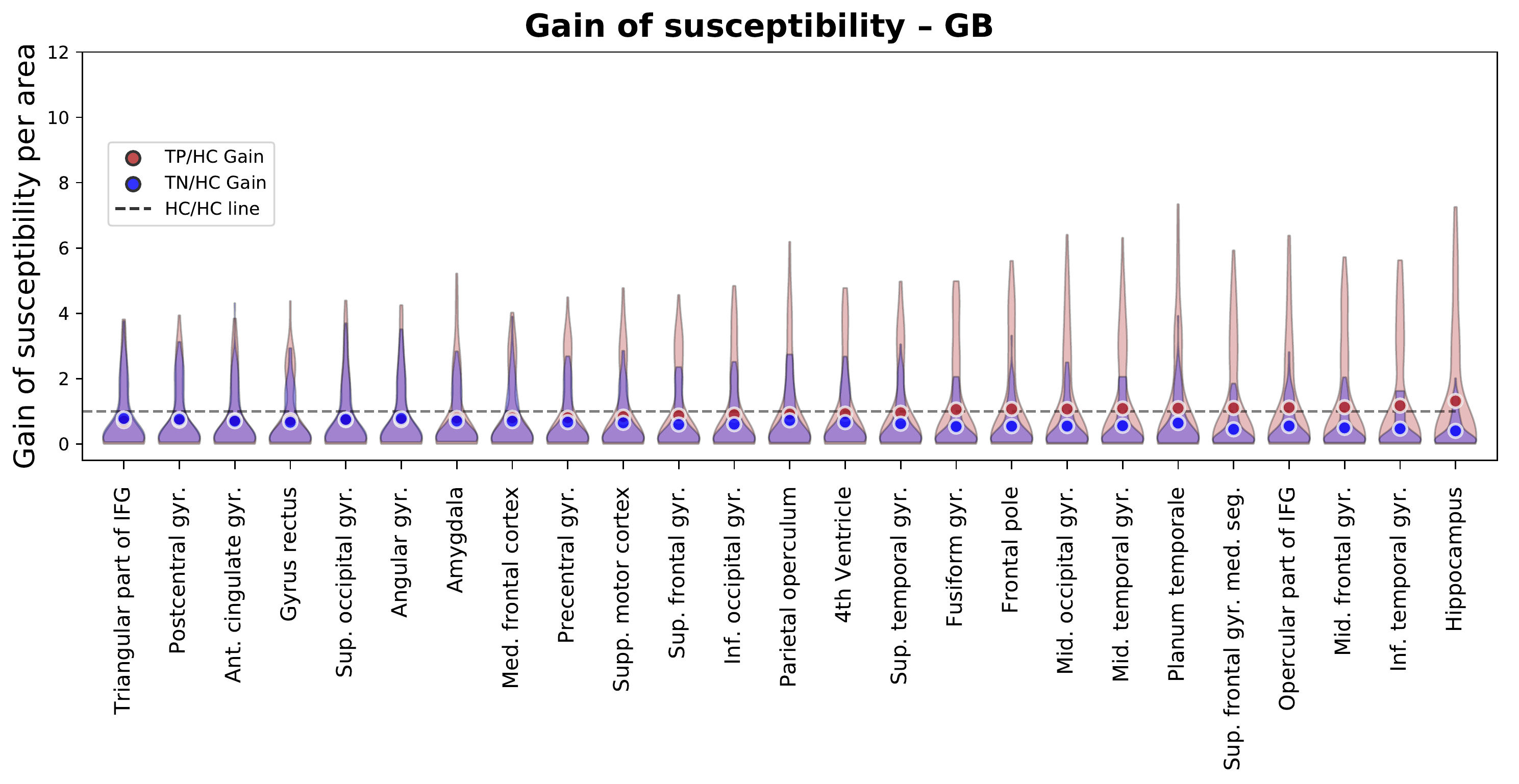}

    \captionof{figure}{
    Gain ofrelevance (LRP, top) and gain of susceptibility (GB, bottom) is shown for different brain areas. The gain per area is defined as the average sum
	of relevance (LRP) or susceptibility (GB) in a given area divided by the average sum in this area
	over all healthy controls (HCs) in the test set. Again, only the top 25 most important areas under this metric are shown for LRP and GB respectively. To provide an estimate of gain in correctly classified subjects, we show here the mean and density estimations only for true positive (TP) and true negative (TN) classifications.
	 As an additional 
	visual aid, the identity gain (gain of 1) is shown as a dashed line.
	}
	\label{fig:ratio_evidence}
\end{figure*}

In section \ref{sec:GP_LRP}, we compare the heatmaps generated by GB and LRP qualitatively with respect to different $\beta$ values and different sets of data (AD, HC, true positives, false positives etc., see Figs.\ \ref{fig:LRP-GB-AD_comparison_betas}-\ref{fig:LRP-GB-cases}). In section \ref{sec:metrics}, we quantitatively compare the heatmaps with respect to the different atlas-based importance metrics (see Figs.\ \ref{fig:abs_evdc}-\ref{fig:intersections}). 
In section \ref{sec:indvd_heat}, we present and discuss the LRP heatmaps of two individual patients (see Fig. \ref{fig:idv_heatmaps}) and investigate the association between LRP relevance scores and hippocampal volume as one of the neurobiological key markers of AD (see Fig. \ref{fig:corr-LRP-GB-Hippo}).

\subsection{Average heatmap comparison}
\label{sec:GP_LRP}
In Fig.\  \ref{fig:LRP-GB-AD_comparison_betas}, we show the average heatmaps for  AD patients and HCs, separately for LRP with different $\beta$ values ($\beta=0,0.5,1$) and GB. The AD pattern between LRP and GB is relatively similar, which is reasonable because all heatmaps are extracted from the same CNN model. However, whereas GB heatmaps are very susceptible for both AD and HCs, LRP heatmaps show much more relevance in AD patients than HCs. This indicates that LRP heatmaps might be more valuable in assessing why a certain person has been classified as AD patient as opposed to which voxels should be changed to increase the likelihood for AD diagnosis. 
Concerning the different $\beta$ values, it is noted that the heatmaps look qualitatively similar, but that sparseness increases with higher $\beta$ values (which is due to a larger effect of inhibitory contributions, see also \citet{binderbeta}). 
Since $\beta$ values close to $0$ focus on positive AD contributions and are thus clinically better interpretable, we focus on $\beta=0$ in the remaining analyses.

In Fig.\  \ref{fig:LRP-GB-cases}, we show the average heatmaps for the distinct classification cases (true positives, false positives etc.), separately for LRP ($\beta=0$) and GB. In particular, the false positives lead to an interesting insight: For LRP, the false positives exhibit less relevance than the true positives, but generally in similar areas. This could indicate that in these
    patients structures that are correlated with AD were found, albeit that overall the positive contribution was less compelling than
    for true AD patients. 
    For GB on the other hand, the false classifications (mostly false positives, but also false negatives) seem to exhibit the highest gradient values of all cases. This exemplifies well what 
    GB truly measures: in the case of false positives (and negatives), the network might be `unsure' and more easily influenced to change its decision; the outcome is unstable. 
    The highlighted areas that
    could change the outcome are very broadly distributed and need not necessarily represent 
    areas with positive contributions for AD.

\subsection{Atlas-based importance metrics}
\label{sec:metrics}
In Fig.\ \ref{fig:abs_evdc}, we show the sum of AD importance per area, separately for LRP ($\beta=0$) and GB. Although this metric seems to be dominated by the size of the respective brain area, one important qualitative difference between LRP and GB is visible:
in the LRP results, the mean importance values per area are consistently much higher for AD patients than for HCs. For GB, this clear split is not present; moreover, the average sum of gradients in several brain regions, including the cerebral white matter and cerebellum, is even higher for HC than for AD. This exemplifies well that the heatmaps for GB cannot directly be interpreted as showing the relevance for AD classification, but instead show the sensitivity of the outcome to certain areas, which does not have to be AD or HC specific. As the absolute sum of importance correlates with the size of the respective brain area, the following metrics, in which we controlled for the brain area size, might be better interpretable.

\begin{figure*}[p]
    \vspace{0cm}
    \centering
  
  \includegraphics[clip, trim=0cm 0cm 0cm 0cm,page=1,
  width=0.95\textwidth]{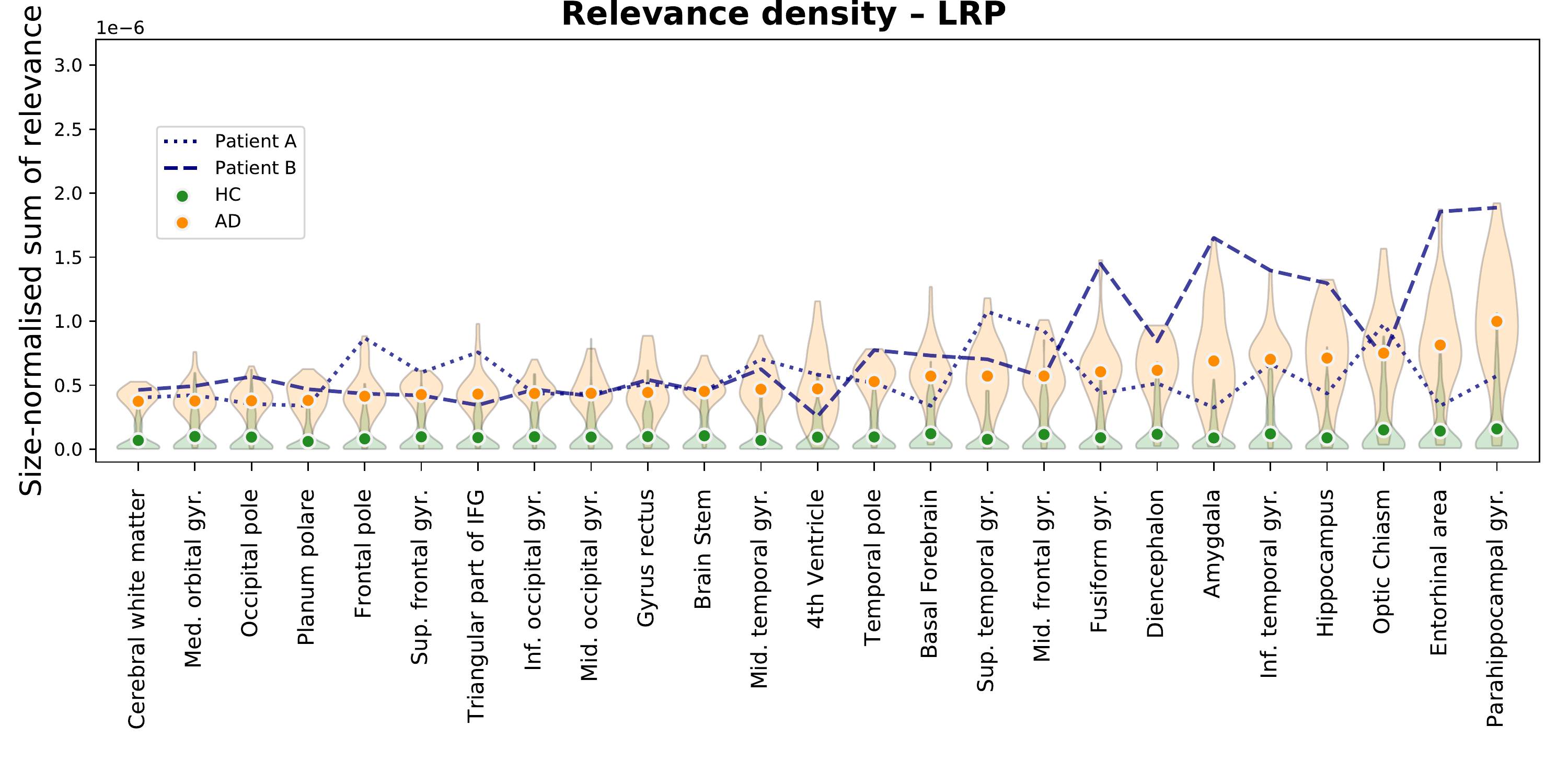}
  \includegraphics[clip, trim=0cm 0cm 0cm 0cm,page=1,
  width=0.95\textwidth]{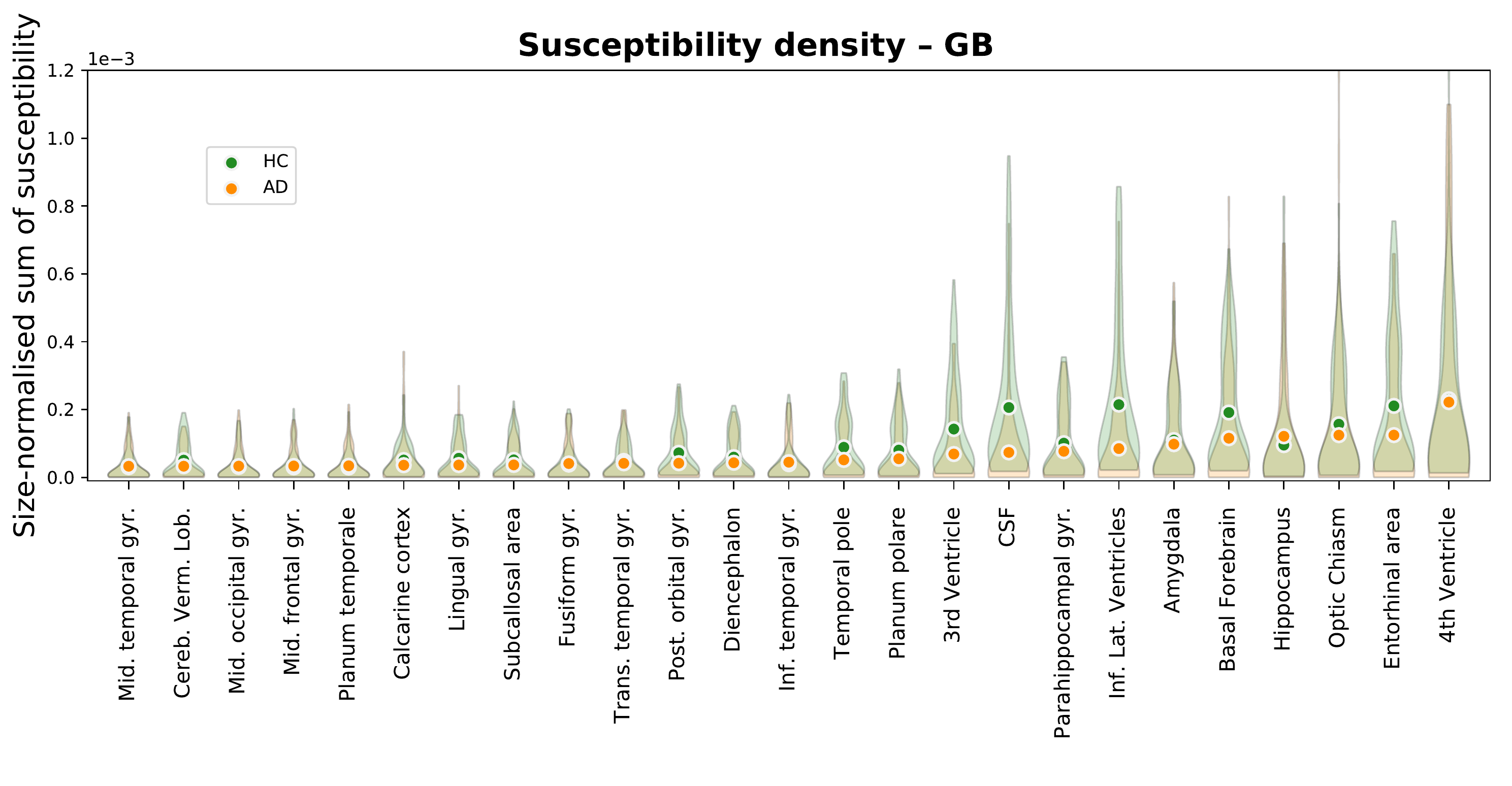}
 
    \captionof{figure}{
	Size-normalized relevance (LRP, top) and size-normalized susceptibility (GB, bottom) is shown for different brain areas. 
	Only the top 25 most important areas under this metric are shown for LRP and GB respectively. We show the average density for all AD patients (orange circles) and all healthy controls (HCs, green circles) in the test set along with a density estimation of the distribution of values per area (orange and green shaded area for AD and HCs respectively).
	    Moreover, two patients were selected to emphasize the diversity in 
	    relevance distributions for LRP; the patients were selected as those with the 
	    highest cosine distance in the relevance-density space of the 25 areas between each other among those patients that were  classified as 
	    AD with a class score $>$90\%.
	    }
	\label{fig:size_norm_evdc}
    \end{figure*}

\begin{figure*}[p]
        \vspace{0cm}
        \centering
      \includegraphics[clip, trim=0cm 0cm 0cm 0cm,page=1,
      width=0.95\textwidth]{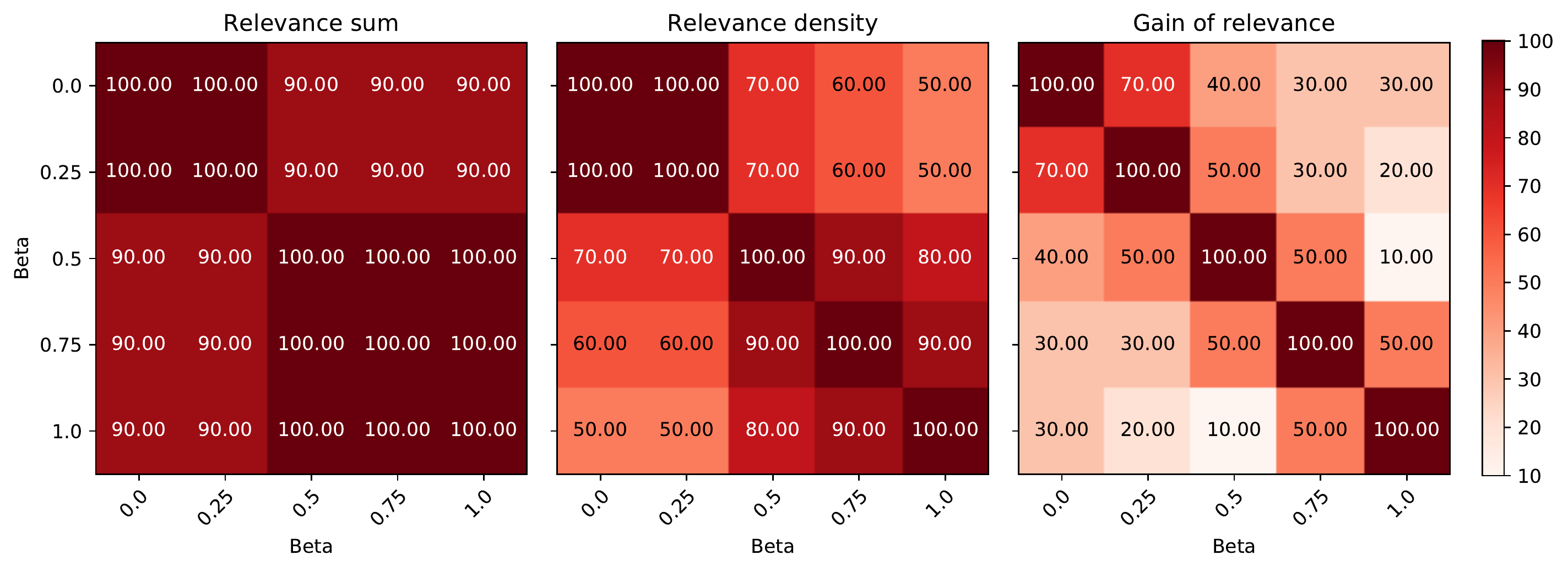}

        \captionof{figure}{ Comparison of the effect of different $\beta$ values on the regional ordering in Figs. \ref{fig:abs_evdc}, \ref{fig:size_norm_evdc} and \ref{fig:ratio_evidence}. The intersection between the top 10 regions of the three metrics is shown for different LRP $\beta$ values in \%.
        }
    	\label{fig:intersections}
    \end{figure*}

\begin{figure*}[p]
    \vspace{0cm}
    \centering
        \includegraphics[clip, trim=0cm 0cm 0cm 0cm,page=1,
        width=.8\textwidth]{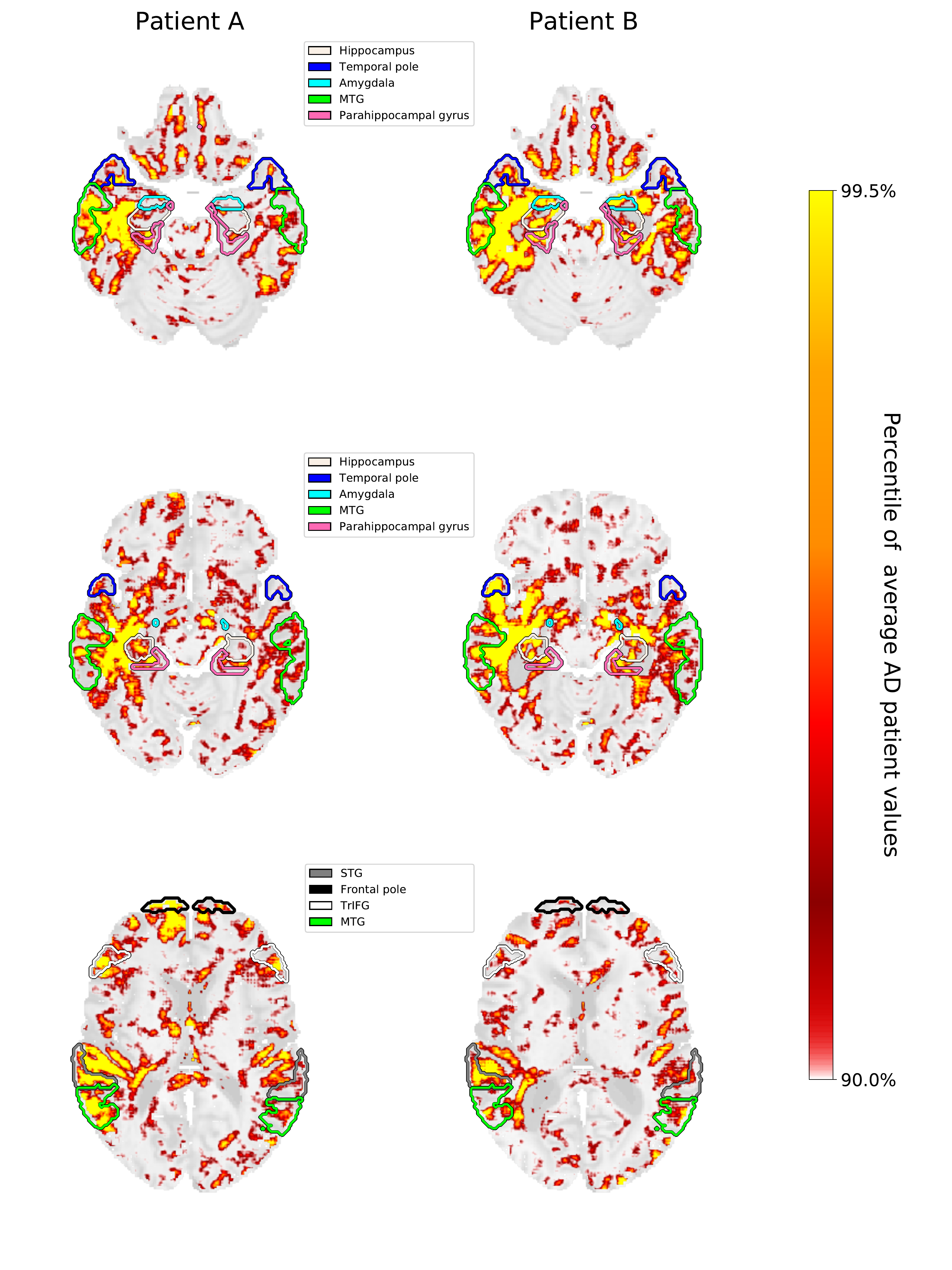} \\
    \captionof{figure}{
    Three brain slices are shown for patient A and patient B, whose individual slopes in relevance density have been shown in Figure \ref{fig:size_norm_evdc}.
	The highlighted areas are the hippocampus, temporal pole, amygdala, parahippocampal gyrus, medial temporal gyrus (MTG), superior temporal gyrus (STG), triangular part of the inferior frontal gyrus (TrIFG) and frontal pole.
	The scale for the heatmap is chosen relative to the average AD patient heatmap. Hence, values in the individual patients that are higher than the 90th percentile and lower than the 99.5th percentile are linearly color-coded as shown on the scale. Values below (above) these numbers are transparent (yellow).
	}
	\label{fig:idv_heatmaps}
    \end{figure*}

In Fig.\ \ref{fig:size_norm_evdc}, the total sum of importance is normalized by the size of the respective brain area. Here, the aforementioned difference in the distributions between HCs and AD patients becomes
even more apparent: while the distributions are very heavily overlapping 
for GB, this is not the case for LRP. Notably, the variance in the AD distributions is much higher in the AD case than in the HC case. 
This could indicate that the network has learned to differentiate between subtypes of AD and bases its decision on different structural elements for 
different patients; the existence of different subtypes of AD has been investigated in recent work, see for example \citep{park2017, ferreira2017distinct}.
In contrast, for HCs the relevance density is consistently very low. As an example of the diversity in importance assessments according
to this metric, we added the `individual fingerprints' of two AD patients to Fig. \ref{fig:size_norm_evdc}; 
for these patients the individual heatmaps will be compared in section \ref{sec:indvd_heat} and Fig. \ref{fig:idv_heatmaps}.

In Fig.\ \ref{fig:ratio_evidence}, the results for the gain metric for different cases -- true positives and true negatives -- are visualized. This metric allows for plotting the LRP and the GB results on the same scale and emphasizes once again the stronger distinction between AD patients and HCs under the LRP algorithm. Most gain for LRP has been found in areas of temporal lobe including transversal temporal gyrus, hippocampus, planum temporale and amygdala.

In Fig.\ \ref{fig:intersections}, we compare the regional overlap of the top 10 regions between the $\beta$ values $0,0.25,0.5,0.75,1$, separately for the three importance metrics. It is shown that (1) the regional overlap is strongest for relevance sum followed by relevance density and relatively unstable for gain of relevance especially for large and more distant $\beta$ values and (2) the regional overlap is -- as expected -- stronger for neighbouring $\beta$ values. The instability of the gain metric for higher $\beta$ values is probably due to the associated sparsity leading to very low relevance scores for HCs (which might -- in some cases -- inflate the gain metric).

\subsection{Individual heatmaps - fingerprinting and neurobiological relevance}
\label{sec:indvd_heat}

Since the LRP heatmaps take into account the individual filter activations and therefore highlight positive contributions to the class score of AD, they might serve as `individual fingerprints' in a diagnostic tool. 
In Fig.\ \ref{fig:idv_heatmaps}, we show several slices of the relevance heatmaps for two patients in order
to highlight the diversity in those heatmaps. The two patients were selected from the test set as those with the 
highest cosine distance in the relevance-density space between each other among those patients that were classified as 
AD with a class score $>90\%$ (their individual trajectories of region-wise relevance are shown in Fig. \ref{fig:size_norm_evdc}). It can be seen that the areas, which mainly contributed to the network decision, are rather different for the two patients. 
For one patient (patient B), the class score of the network is heavily influenced by areas of the temporal lobe, such as parahippocampal gyrus, entorhinal area, hippocampus, inferior temporal gyrus and amygdala, while for the second patient (patient A), frontal areas, including triangular part of inferior frontal gyrus, superior frontal gyrus and frontal pole, in addition to superior temporal gyrus seem to be most informative. 

\begin{figure*}[p]
    \vspace{0cm}
    \centering
        \includegraphics[clip, trim=0cm 0cm 0cm 0cm,page=1,
        width=.8\textwidth]{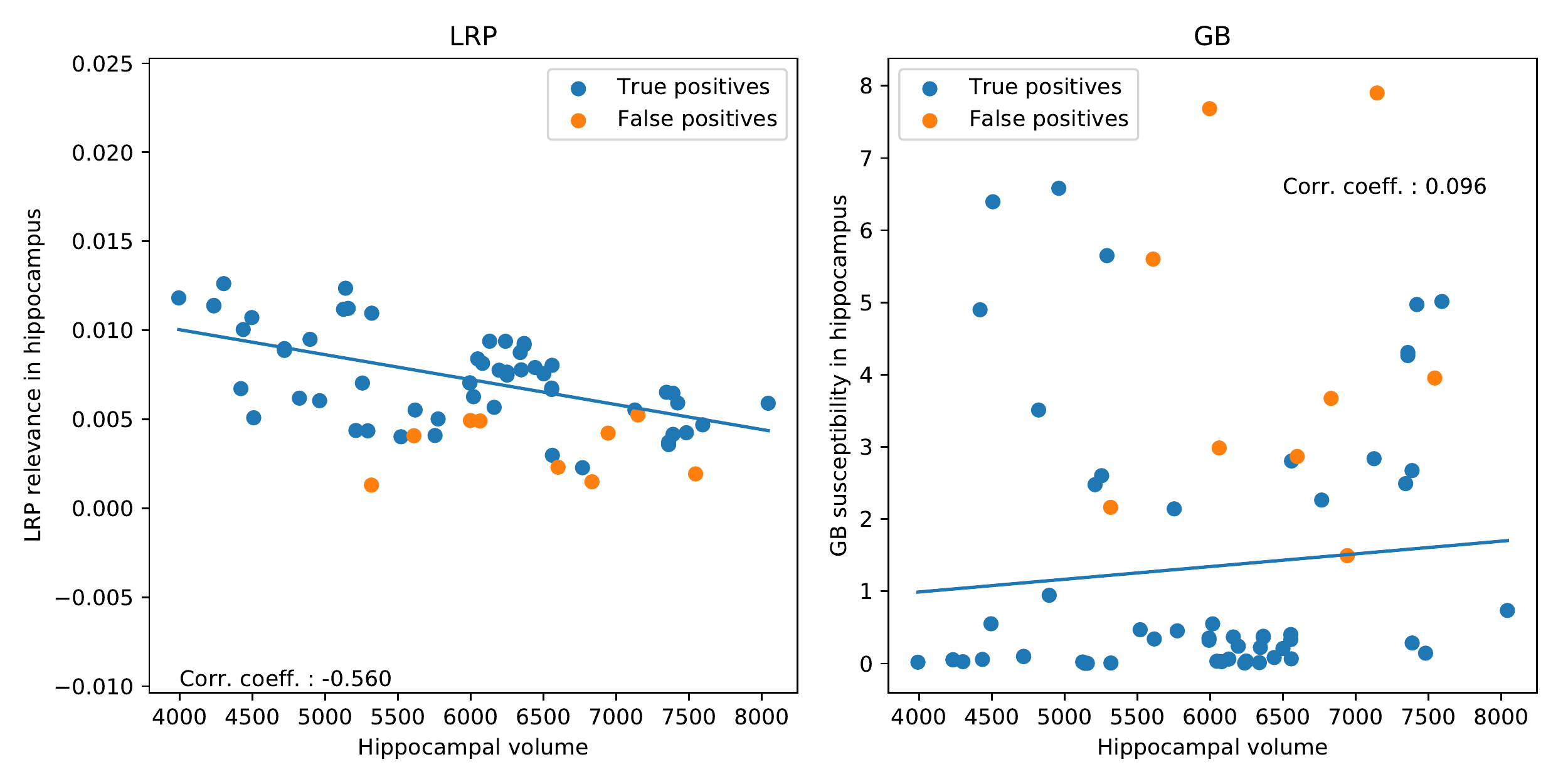} \\
    \captionof{figure}{Correlation between hippocampal volume and LRP relevance / GB susceptibility in hippocampus for correctly classified AD patients (true positives; left: LRP, right: GB). For illustration, we show additionally the false positive classifications. 
	}
	\label{fig:corr-LRP-GB-Hippo}
    \end{figure*}

To investigate whether higher importance scores correspond to stronger anatomical deviations (e.g. atrophy) in correctly classified AD patients (true positives), we performed a correlation analysis between hippocampal volume and LRP relevance / GB susceptibility scores (see Fig.\ \ref{fig:corr-LRP-GB-Hippo}). We show that the LRP relevance score ($\beta=0$) in the hippocampus is significantly (negatively) correlated with hippocampal volume (-$0.560, p<10^{-3}$, permutation test), whereas the GB score is not ($0.096, p = 0.77$). To rule out that false positives are outliers in terms of association between hippocampal volume and LRP relevance, we included them in Fig.\ \ref{fig:corr-LRP-GB-Hippo}. Interestingly, for larger $\beta$ values the correlation tends to decrease ($-0.560,-0.562,-0.525, -0.457,-0.361$ for $\beta=0,0.25,0.5,0.75,1$ respectively) supporting our notion of a higher neurobiological relevance in case of $\beta$ values close to $0$.

\section{Discussion}
\label{sec:discussion}
In this study, we introduced LRP as a powerful method for explaining individual CNN decisions in AD classification. 
After training a CNN to separate AD patients and HCs based on structural MRI data, individual heatmaps -- indicating the importance for each voxel for the respective classification decision -- were produced for the test subjects. 
We analyzed the heatmaps with respect to different classification subgroups (AD patients, HCs, true positives, false positives etc.) and different $\beta$ values.
 The relevance of brain regions contained in the Neuromorphometrics atlas was evaluated using three different importance metrics, namely the sum of importance per area, the size-normalized AD importance, and the gain as ratio between AD and HC importance. 
We demonstrated that LRP-derived heatmaps -- in contrast to GB -- provide (1) high specificity for individuals and (2) little relevance for AD in HCs. Additionally, areas that exhibit a lot of relevance correlate well with what is known from literature. Importantly, these LRP heatmaps were produced without the need for
expert annotations on the presence or absence of biomarkers throughout the learning process. This combination of a simple classification task (AD vs.\ HC) and in-depth network analysis
by LRP might be a promising tool for diagnostics. Additionally, it could allow for discovering new and unknown biomarkers for a variety of diseases and
might help distinguishing subtypes of AD by analyzing the diversity in `relevance hot-spots' across all AD patients.
Furthermore, the size-corrected metrics (`relevance density' and `relevance gain') seem to correlate well with what is known from AD research, indicating that the most discriminating features for classifying an input image as AD can be found in the temporal lobe.  
We therefore think that a well-trained neural network, analyzed by means of the LRP algorithm, can become a useful tool for practitioners and increase the trust in computer-aided diagnoses, as an interpretable explanation of the decision can be produced.

\subsection{Regional specificity of LRP}
We quantitatively evaluated the heatmaps, obtained by either GB or LRP, 
towards different brain areas according to the Neuromorphometrics atlas \citep{bakker2015scalable} by summarizing the importance (AD relevance in case of LRP, susceptibility in case of GB) for each brain area separately. Both types of heatmaps mostly identified regions known to be important in disease progression of AD, such as structures in the medial temporal lobe including hippocampus, amygdala, parahippocampal gyrus, and entorhinal cortex \citep{Du441,Desikan2009, frisoni2010clinical, Weiner2013,  Velayudhan2013, Klein-Koerkamp2014, long2017prediction} as well as frontal and parietal areas \citep{ Casanova2011, Quiroz2013,kilimann2017parallel, park2017, Liu2018a}. For all these regions morphometric changes including global and local atrophy (e.\,g.\,smaller volumes of hippocampus or amygdala, reduced cortical thickness or grey matter density) or deviations in shape  have been shown and related to disease progression and cognitive decline \citep{Desikan2009, frisoni2010clinical,Weiner2013, hidalgo2014regions,long2017prediction, ledig2018structural}. 
These changes seem to be utilized by our CNN framework for making individual predictions and are highlighted in the heatmaps of both LRP and GB. However, the contrast in importance scores between AD patients and HCs is much higher for LRP than GB (in GB, the average heatmaps for AD patients and HCs are quite similar). 
This supports the notion that LRP heatmaps reflect AD-specific relevance, whereas GB emphasizes areas which the network more generally is sensitive to. 
Regarding other structures found to be important in our network, it might be interesting to see if also other neural networks find relevance in these areas and if predictions about finding significant structural changes in these areas might be possible at some point. In this respect, the decisions of such networks can be treated as a `second opinion' and a reciprocal learning process with medical experts might be initiated. 

\subsection{Fingerprinting and neurobiological relevance}
In addition to heatmap differences between AD patients and HCs, we noticed a high variability between the heatmaps of individual AD patients for the LRP method. This variability was not only reflected in a high variance of important scores within regions, but also in individual trajectories (`fingerprints'), which we exemplary depicted for two AD patients, see Fig.\ \ref{fig:idv_heatmaps}. For future work, it might be very interesting to see if these fingerprints reflect different disease stages of AD \citep{Braak1991, Casanova2011} or allow for identifying subtypes of AD, in which brain areas are affected differently \citep{Murray2011, noh2014anatomical, scheltens2016identification, Zhang2016, ferreira2017distinct, park2017}. \citet{Zhang2016}, for example, identified a temporal, a subcortical and a cortical atrophy factor associated with impairment in different cognitive domains.
Another important question is whether the relevance found by the LRP method reflect some true evidence in the sense of biomarkers. By showing that the hippocampal volume is significantly correlated to the LRP relevance scores (but not to the GB susceptibility scores), we argue that LRP -- at least partially -- succeeded here in breaking down the relevance to the level of voxels in a meaningful way. Interestingly, we found higher correlations for lower $\beta$ values speaking for a higher neurobiological relevance of $\beta$ values close to 0. Further studies are needed to more carefully relate LRP relevance measures to other clinical markers of AD including biomarkers and neuropsychological test scores, also in dependency of different CNN models and parameter settings. Moreover, our metrics should be evaluated in patients with mild cognitive impairment (MCI).

\subsection{Related work}
Visualization of deep neural networks is a fairly new research area and different attempts have been made to provide intuitive explanations for neural network decisions. However, there is not yet a state-of-the art visualization method as saliency maps for example have been shown to be misleading \citep{Adebayo2018}. In Alzheimer's research, there are only a couple of studies that looked into different visualization methods based on MRI and/or PET data. Most of these studies either visualized filters and activations of the first or last layer \citep{Sarraf2016, Ding2019, Lu2018a} or used the occlusion method to exclude some parts (e.\,g.\,with a black patch) of the input image and recalculate the classifier output \citep{Korolev2017, esmaeilzadeh2018end, Liu2018a}. Based on visual impression, they found that the networks focus primarily on areas known to be involved in AD, such as hippocampus, amygdala or ventricles, but occasionally also other areas such as thalamus or parietal lobe appear. Importantly, in contrast to our study, they did not quantitatively analyze the data, e.\,g.\,with respect to brain areas contained in an atlas or underlying neurobiological markers. Additionally, they did not compare different visualization methods or looked for inter-individual differences. One study, however, used gradient-weighted classification activation mapping (grad-CAM) and compared it to sensitivity analysis for AD classification \citep{yang2018visual}. They demonstrate that these different visualization methods capture different aspects of the data and show high variability depending e.\,g.\,on the resolution of the convolutional layers. In \citet{rieke2018visualizing}, gradient-based and occlusion methods (standard patch occlusion and brain area occlusion) were qualitatively and quantitatively compared for AD classification. High regional overlaps between the methods, mostly inferior and middle temporal gyrus, were found but for gradient-based methods the importance was more widely distributed. Regarding the LRP method, we are only aware of one application in the neuroimaging field: \citet{Thomas2018} introduce interpretable recurrent networks for decoding cognitive states based on functional MRI data and demonstrate that the LRP method is capable of identifying relevant brain areas for the different tasks and different levels of data granularity. 

\subsection{Limitations}
Although LRP heatmaps seem to be a promising tool for visualizing neural network decisions, we would like to point out several limitations of LRP and other heatmap methods in the context of this study. 

    First, heatmap methods are limited by the lack of a ground truth. Most commonly, heatmaps are qualitatively evaluated based on visual assessment, but there are also studies proposing sanity checks \citep{Adebayo2018} or more objective quality measures such as region perturbation \citep{Samek2015}. In \citet{Lipton:2018}, the interpretability of models has been generally investigated and questioned. In medical research, heatmaps can be qualitatively evaluated based on prior knowledge (e.\,g.\,hippocampus is known to be strongly affected in AD, therefore it seems reasonable to find relevance there). Given that in the specific case of heatmaps for MR images the input space is highly structured, we proposed here additional ways for assessing the quality of explanations by using a brain atlas. Future studies might assess the neurobiological validity by removing presumably important brain areas and re-training the classifier. 
   
    Second, it is largely acknowledged that heatmaps are quite sensitive to the specific algorithms (and its parameters, e.g. the $\beta$ value in case of LRP) used to produce them. However, regarding the $\beta$ values in LRP, we have shown that the heatmaps are relatively robust towards this parameter, only sparsity increases as a function of $\beta$. Additionally, we demonstrated that the regional ordering is relatively stable 
    for relevance sum and density, but unstable for the gain metric -- especially in the case of large and more distant $\beta$ values.
    
   Third, heatmaps just highlight voxels that contributed to a certain classifier decision, but do not allow making a statement about the underlying reasons (e.\,g.\,atrophy or shape differences) or potential interactions between voxels or brain areas. 
    For example, it is difficult to disentangle interactions between different regions 
    (certain patterns in the hippocampus might only be considered as positive evidence if structure Y is found in area Z) 
    nor do we know whether the network developed specific filters for atrophy or the shapes of different structures. 
    Although we found in this study a significant correlation between hippocampal volume and LRP relevance measures, we can not make any claim about causal relationships here. Future studies are necessary to more systematically investigate the relationship between manifested neurobiological markers and LRP explanations.
    
    Fourth, heatmaps strongly depend on the type and quality of the classifier, whose decisions are sought to be explained. Therefore, each heatmap should be read as 
    an indication of where the specific network model sees evidence. For badly trained networks, this does not have to correlate at all with the presence of actual biomarkers.
    Nevertheless, the better the classifier,
    the more likely it becomes that 
    the classifier uses meaningful patterns as a basis for its decision and that the heatmaps correlate with `true' evidence for AD. 
    However, heatmaps are also useful in cases, where classification performance is low or sample size is rather small, e.\,g.\,for better understanding if the classifier picks up relevant or irrelevant features (e.\,g.\,noise or imaging artifacts) and if there are any biases present in the data set \citep{Lapuschkin_2016_CVPR, Montavon2018}. It would be very interesting to investigate how the heatmaps change for different networks, as 
those which yield stronger classification results should also base their decisions on \textit{better} `evidence'.
    
    And finally, it should be stressed that when we refer to brain areas throughout this work, we refer to the location that the areas are assigned in the brain atlas 
    and not to the individual anatomical structures of any patient. Due to inter-individual differences, the match between the atlas and the individual patient's 
    anatomical realities will not be perfect; this is most likely further aggravated by the presence of atrophy in AD patients.

\section{Conclusion}

In conclusion, we introduced the LRP method for explaining individual CNN decisions in MRI-based AD diagnosis. In contrast to GB, LRP heatmaps can be interpreted as providing individual AD relevance (``What speaks for AD in this particular subject?") as opposed to a general susceptibility for small variations in the input data. Additionally, we provided a framework and specific metrics (i.\ e.\ `relevance density' and `relevance gain') to quantitatively compare heatmaps between different groups, brain areas or methods. We demonstrated that these metrics correlate well with clinical findings in AD, but also vary strongly between AD patients. By this, the LRP method might be very useful in a clinical setting for a case-by-case evaluation. However, we would like to point out that (1) our metrics should be evaluated in different network architectures and (2) other (individual) brain atlases might be used for the evaluation of regions. Future studies should evaluate the LRP method on patients with mild-cognitive impairment (MCI) and relate findings to known biomarkers in AD. We are convinced that our framework might also be very useful for other disease classification studies in helping to understand individual network decisions.

\section*{Conflict of Interest Statement}

The authors declare that the research was conducted in the absence of any commercial or financial relationships that could be construed as a potential conflict of interest.

\section*{Author Contributions}
MB, FE, MW and KR designed the study. MB and FE engineered the software. MB and FE analyzed the data. MB, FE and KR wrote the paper. 

\section*{Funding}
We acknowledge support from the German Research Foundation (DFG, 389563835), the Manfred and Ursula-Müller Stiftung and Charité – Universitätsmedizin Berlin (Rahel-Hirsch scholarship and Open Access Publication Fund).

\section*{Acknowledgments}
Data  collection  and  sharing  for  this  project  was  funded  by  the  Alzheimer's  Disease Neuroimaging  Initiative  (ADNI)  (National  Institutes  of  Health  Grant  U01  AG024904)  and DOD  ADNI  (Department  of  Defense  award  number  W81XWH-12-2-0012). ADNI  is  funded by  the  National Institute  on  Aging,  the  National  Institute  of  Biomedical  Imaging  and Bioengineering, and through generous contributions from the following: AbbVie, Alzheimer’s Association;  Alzheimer’s  Drug  Discovery  Foundation;  Araclon  Biotech;  BioClinica,  Inc.; Biogen; Bristol-Myers   Squibb   Company;CereSpir,   Inc.;Cogstate;Eisai   Inc.;   Elan Pharmaceuticals,  Inc.;  Eli  Lilly  and  Company;  EuroImmun;  F.  Hoffmann-La  Roche  Ltd  and its  affiliated  company  Genentech, Inc.;  Fujirebio;  GE  Healthcare;  IXICO  Ltd.;  Janssen Alzheimer    Immunotherapy   Research   \&   Development,   LLC.;   Johnson   \&   Johnson Pharmaceutical  Research  \&  Development  LLC.;Lumosity;Lundbeck;Merck  \&  Co.,  Inc.; Meso  Scale  Diagnostics,  LLC.;NeuroRx  Research;  Neurotrack  Technologies;Novartis Pharmaceuticals Corporation; Pfizer Inc.; Piramal Imaging;Servier; Takeda Pharmaceutical Company;  and  Transition  Therapeutics.The  Canadian  Institutes  of  Health  Research  is providing  funds  to  support  ADNI  clinical  sites  in  Canada.  Private  sector  contributions  are facilitated by the Foundation for the National Institutes of Health (\url{www.fnih.org}). The grantee organization is the Northern California Institute for Research and Education, and the study is coordinated by the Alzheimer’s Therapeutic Research Institute at the University of Southern California.  ADNI  data  are  disseminated  by  the  Laboratory  for  Neuro  Imaging  at  the University of Southern California.

\section*{Data Availability Statement}
The ADNI data set is for researchers publicly available at \url{http://adni.loni.usc.edu/}. The code is available at \url{https://www.github.com/moboehle}.

\bibliographystyle{frontiersinSCNS_ENG_HUMS} 
\bibliography{frontiers}

\end{document}